# Application of Reverse Engineering and Rapid Prototyping for Reconstruction of Human Mandible

## A Thesis

Submitted to the Department of Production Engineering and Metallurgy of the University of Technology in a Partial Fulfillment of the Requirement for the Degree of Master of Science in Industrial Engineering

By
Nabeel Ismail Allawi
BSc. Industrial Engineering 2006

Supervised By
Dr. Amjad B. Abdulghafour

2020

ص

ح

# Acknowledgments

At first, "Thank God" for helping me complete this work.

I would like to express my deep appreciation and gratitude to the supervisors of Dr. Amjad Barzan Abdul Ghafoor, who plays a great role in providing the basis for this research, for his guidance, advice, and encouragement. I would like to thank the Chairman and other members of the Department of Production Engineering and metallurgy. My deep thanks to all the people who helped me achieve this work. I would also like to thank my family for the support they provided to me throughout my life and especially my mother, and in particular, I must confess my wife, who without her love, encouragement, and help in mobilization, I would not have finished this work.

# Dedication

*To those who wish me all the best and good luck in science, my father (God rest his soul), my mother, my brothers and sisters, and all my friends.*

*To who helped me to achieve my scientific career, my dear wife.*

*To my children, to guide to the path of knowledge.*

# List of Contents

# List of Figures

# List of Tables

# List of Abbreviations

| 2D/3D | Two- Dimensional/ Three- Dimensional |
|---|---|
| AM | Additive Manufacturing |
| CAD/CAM | Computer-Aided Design/Computer-Aided Manufacturing |
| CAE | Computer-Aided Engineering |
| CBCT | Cone Beam Computerized Tomography |
| CT | Computerized Tomography |
| CMM | Coordinate Measuring Machine |
| DICOM | Digital Imaging and Communications in Medicine |
| FDM | Fused Deposition Model |
| FEA | Finite Element Analysis |
| HU | Hounsfield Unit |
| IGES | International Graphics Exchange System |
| LOM | Laminated Object Manufacture |
| MRI | Magnetic Resonance Imaging |
| RE | Reverse Engineering |
| RP/RPM | Rapid Prototyping/ Rapid Prototyping Manufacturing |
| SLA | Stereolithography |
| SLS | Selective Laser Sintering |
| STL | Standard Triangle Language |
| ZOMC | Zygomatico-Orbito-maxillary complex |

# Abstract

Reconstruction of the mandible is one of the most common challenges facing maxillofacial surgeons. The mandible plays a major role in supporting the teeth inside the mouth.

This work aims to develop a proposed methodology for improving reconstructive surgery by using the simulation of a mandibular defect using imaging, design and fabrication techniques for a custom mandible. The combination of these technologies provides a powerful way to improve and implement the implant process through the design and fabrication of medical models.

This work introduces a methodology that the capabilities of Reverse Engineering (RE), Computer-Aided Design (CAD), and Rapid Prototyping Technology (RP) for the reconstruction of the mandible and representing the surgery. Patient examination using a high-resolution technique represented by Cone-beam computed tomography (CBCT), after which a representative digital model of the patient is created to assist in the design process using the 3DSlicer software The patient's implant designed (Defect part) and analyzed using Solidwork.3D models are assembled and implant simulations are performed by (Meshmixer) software, Stereolithography technology (SLA) used as 3D printing technology to fabricate the mandible model the resin.

The results of this work show that procedures for mandible reconstruction can be successfully used using the integration of RE / CAD / RP technologies. This integration will support the acceptable symmetry of the face that will be restored by assisting surgeons in planning reconstruction. This work demonstrates the importance of the CAD system in resolving patient damage requirements directly from medical imaging data, as well as the efficiency of RP techniques used in converting 3D model designs into implant models.

# CHAPTER ONE

# Introduction

# CHAPTER ONE

## Introduction

### 1.1 Additive Manufacturing

Additive Manufacturing (AM) processes, which are also known as Rapid Prototyping (RP), refer to an evolutionary type of fabrication that utilizes a 3D computer-aided design CAD file and slices it to different thicknesses [1]. A computer uses the sliced files as the geometry of each layer and orders the fabrication setup to deposit a layer regarding that geometry. The layer-by-layer deposition continues to the last layer in order to fabricate a complete 3D component. Various deposition methods are working on a different basis. However, these processes are similar in the thermal, chemical and mechanical ways they fabricate parts. The most common processes in the medical field are Stereolithography (SLA) and Fused Deposition Modelling (FDM). AM processes can be categorized into solid, liquid and powder-based types that as comes from their names, they are working with solid, liquid and powder feedstock [2].

Traditionally, the AM process begins with the generation of a three dimensional (3D) model by using (CAD) software. Typically, as a Standard Tessellation Language (STL) file, the CAD-based 3D model has been saved which is a triangulated representation of the part. The software then slices the data file into unique layers, which are sent as an instruction to the AM tool [3].

Therefore, the methodology proposed in this work will produce anatomical mandible, which is the most important objective of improving surgery. This methodology allows the surgeon to have a broader view of the surgical area and enables him to pre-plan the surgery. This is better than taking risks and planning surgery within the operating room.

## 1.2 Statement of the Problem

Today, Medical applications are facing great challenges concerning replacements of bones, organs, and tissues that are not readily available for surgery. In addition to the lack of a clear vision of the anatomical area for the purpose of proper planning before surgery Although rapid prototyping is used in this field Current research has not simplified the method of integration between reverse engineering and rapid prototyping environment. This work tries to use RE and RP technology to develop an integrated environment that helps designer needs and professionals identify the tasks that are required in the clinical world.

## 1.3 Aim of Research

1- This work aims to provide a method for designing and manufacturing a medical model using CAD and RP techniques and implementing them in oral and maxillofacial surgery. For this purpose, an environment has been developed between imaging techniques, design, and manufacturing.

2- Study the possibility of designing the reconstruction of the mandible and restore the patient's functional performance. Using software such as 3DSlicer 4.10.2, SolidWorks 2015, Meshmixer 3.5 and UltimakerCura Cura 4.4.1.

## 1.4 Importance of Research

In this study, an advanced (innovative) method was presented in the applications of oral and maxillofacial surgery the design of the implant to the patient and improves the outside appearance. CAD system and fabrication methods have several advantages, as they are well suited to the anatomical area, reduce surgery time, and produce a more appropriate cosmetic appearance. On the other hand, manual methods need more time in the design process while their success depends largely on the skill of the surgeon.

## 1.5 Thesis Layout

This section describes all the research chapters and the content of the main thesis as follows:

**Chapter 2:** Contains a background of RP & RE and principles of the RE, an illustration of the RP principles and method RP in the medical field with the type technique.

**Chapter 3:** This chapter includes the methodologies that researchers have found relevant studies as well as a summary of the research.

**Chapter 4**: This chapter includes the proposed new methodology for mandible reconstruction and virtual surgery while discussing the results.

**Chapter 5:** This chapter ensures the implementation of the proposed method for reconstruction of the mandible. Methods and techniques used to build medical models and programs are used in the design and improvement of implants. In addition to virtual surgery procedures and how to perform them.

**Chapter 6:** This chapter includes the conclusion and recommendations.

# CHAPTER TWO

# Reverse Engineering and Rapid Prototyping

# CHAPTER TWO

# Reverse Engineering and Rapid Prototyping

## 2.1 Introduction

Engineering is a technological principle in designing, manufacturing, building and maintaining products, systems, and structures. At different levels, there are two types of engineering: progressive frontal engineering and reverse engineering. "Front-end engineering" is the most traditional method of technological development for a particular product. Technologically business owners are developing a particular product by applying and improving engineering concepts. On the contrary, RE begins with the final product, where basic engineering concepts formed from design analysis and interrelations between components. Frontal engineering is the traditional process of a progressive transition from high abstraction level and designs to actual system implementation and final product formation. Sometimes, there is a physical part without technical details, such as graphics or without geometric data, such as thermal, electrical and engineering dimensions. "Replicating an existing component, subgroup, or current product without the help of graphics, documents, or the computer model as RE". The design of the new products in the CAD system is being to develop from the first stage depending on the RE that details the new product details, different product stages can be developed [4],[5].

RE has become an effective way to produce 3D models for a natural part that exists with the help of CAD and other software. The thing can be measured using 3D scanning technology such as laser scanners. The measured data lack the topological information and processed into a usable form such as the computer-aided design model. 3D model using RP based on 3D CAD model data. The first RP method introduced for the first time in the 1980s in the field of engineering for manufacturing a solid 3D model based on CAD evidence using a special device called Stereolithography. A method where the layers of polymer material are very sensitive to the laser beam [3].

RE and medical image-based modeling technologies allow the construction of 3D models of anatomical structures of the human body based on anatomical information from scanning data such as computerized tomography (CT), magnetic resonance imaging (MRI), and laser (or structured light) scanning. RE and CT/MRI scanners are, respectively, used for scanning data acquisition of the external and internal geometry of anatomical structures. Physical models of anatomical structures can be fabricated directly from 3D digital data by RP technology. With the use of RP and rapid tooling technology, nowadays, we can fabricate complex 3D physical objects in a wide range of materials and sizes, from plastics and metals to biocompatible ones, and from big models to microstructures [6]. Medical RP technology is a multi-discipline area, which applies biomedical modeling and RP to develop medical applications. It involves human resources from the fields of RE, design and manufacturing, biomaterials and medicine. Medical RP has played an important role in diagnosis and treatment; especially in preoperative planning, design, and manufacturing [7].

## 2.2 Reverse Engineering (RE)

The process of duplicating an existing component, subassembly, or product, without the aid of drawings, documentation, or computer model is known as RE. In industrial practice, new product designs are usually created in a CAD system from the beginning. Geometric models are digital and may be used in further computer-aided stages of product development. There are cases, however, when product geometry must be created from an existing physical object, e.g. to re-engineer a design of a product without any technical documentation or to transform an artist's view into industrial design. This is the job for RE, where the physical artifact is digitized into a computer model. The input to RE technologies is a material object, and the output is a CAD model. This can be achieved using contact and non-contact technologies. Among contact technologies most widely used are measurements with a Coordinate Measuring Machine (CMM), among non-contact methods laser scanning and computed tomography. Contact and non-contact methods using visible light allow for reconstruction of geometry of outer surfaces of objects. Models representing objects with uniform internal structure may be recreated this is a sufficient approximation for some applications [1],[8].

## 2.3 Principles of Reverse Engineering

RE based on several principles for successful application as follows:

1- Organizing the design and development of all stages of product operations.

2- Define and review all the stages in which the production of a particular product goes into the light of the proposed design.

3- Feasibility study of product design and manufacturing processes.

4- Achieve less response time and delivery to the desired product.

5- To apply the reference comparison method, continuously [9].

## 2.4 Reverse Engineering Processes

RE begins by specifying the project objective, the appropriate way to determine the geometry of the required system and its parts, the appropriate accuracy of the cutting dimensions, and the method of using the results. As in Figure (2-1) [10].

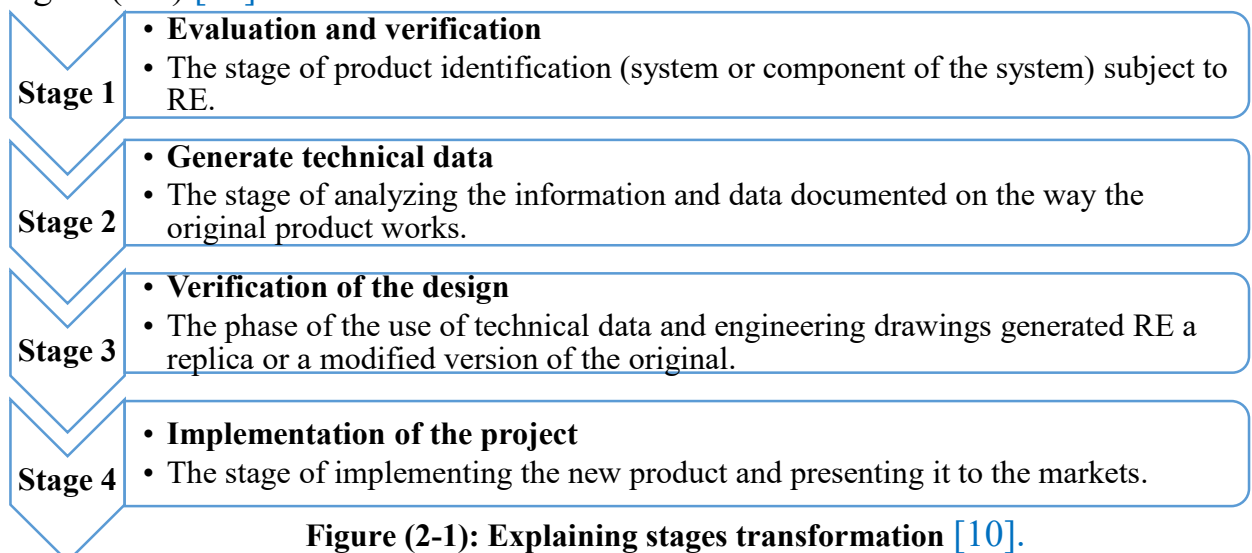

Figure (2-1): Explaining stages transformation [10].

After the success of the implementation of the prototype, the application of all peripheral tests, performance tests, and ensure the integrity of the performance of the product compared to the performance of the original model can introduce the new product in the market. The system is a competitive design in the market; it depends on the original product innovation in terms of technical specifications, efficiency, and technical age, with improvements, and the use of technologies that enable the new product to compete in terms of quality and financial value.

## 2.5 Rapid Prototyping

A new generation of machines has emerged, with different techniques and materials, allowing innovation in order to obtain a prototype of a model or template digitally created in the CAD-3D system, accurately and quickly. These technologies are known as RP, which is quite different from traditional technologies, because it allows you to get the final physical parts automatically, of any shape and dimensions, with a high level of complexity and detail. RP is a process that uses a range of techniques to manufacture 3D plastic, metal or ceramic materials also known as "additive technology" using CAD design. It also offers greater speed and lower cost of obtaining prototypes, which means greater support in the field of industrial design.   Continuous improvement and advanced processing of digital data with specialized software allow RP in a wide range of areas. Such advances, such as so-called biometric models, which are accurate models in 3D, have enriched medical and medical engineering services in particular. It achieved with acrylic resin, which provides a replica of the patient's area. "If the picture worth's a thousand words," the prototype is a thousand images, "and Design has used this concept since before the existence of the prototyping machines [11].

Through the prototype, multiple variables, such as aerodynamics, can be a stud to check the brightness of your body and be primarily able to see and study general acceptance, among many other features required in these models.

Similarly, by using these real sizes, RP, biomedical engineering and engineering could study symmetry, projection, size, defects, distortions and other problems in a more detailed way. Biological models, therefore, are very useful for simulating and preparing complex surgical interventions. Because there is a model for the area where patients' illnesses are presented in advance, a health professional can analyze and assess the complexity of the problem they face before the intervention. In this way, the duration of the intervention is minimized, the patient's risk is reduced to a minimum and the results are better [12].

The information about each level is sent electronically layer by layer to the RP machine and the layers are processed sequentially until the part to be manufactured is complete. The RP method is whether it is sequential, class, or printed. The RP process uses the following steps to create prototypes [13]:

1. Build a CAD model for design.
2. Save the CAD form to the STL file format.
3. Slice the STL file into 2D cross-sectional layers.
4. Build the prototype.

## 2.6 Principles of Rapid Prototyping

Factors influencing the basic approach of various RP techniques are to design the physical part that represents an object or model using a CAD / CAM system that is constructed as a closed surface. Transform the solid or surface model created in a format is called the STL file format. In addition, the specialized computer program analyzes the STL file, which represents the 3D model to be manufactured into segments, "slides". The slides are reconstructed sequentially through the hardening of liquids or powder layer by layer and then combined to form a 3D model. The RP development process is carried out in four main areas as shown in Figure (2-2) [14].

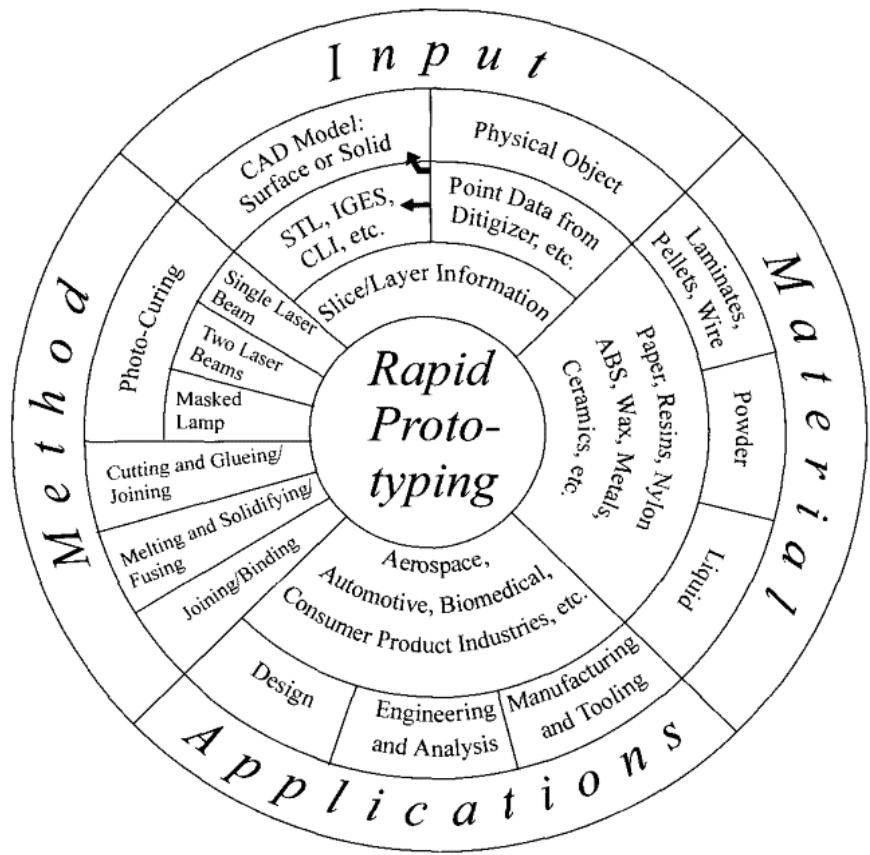

**Figure (2-2): RP Wheel of RP process [10]**

## 1- Inputs

The entry refers to the required data created to describe the physical object using the computer (the solid form or surface) [10].

## 2- Method

The number of manufacturers up than producing RP systems and the number is increasing rapidly at present, the way each company works can be classified into the following classifications [10]:

- Image- Treatment
- Cutting and Gluing
- Melting and Fusing
- Joining and Binding

### 3- Materials

Materials used in several cases and include solid, liquid or powder. In the solid-state, come in different photos such as pellets or wires. Materials can come in the form of paper, nylon, wax, resins, metals, and ceramics. One of the most appropriate methods for classifying these processes are:

- Liquid-Based
- Solid Based
- Powder Based

### 4- Applications

Professional applications can be grouped broadly in the following areas:

- Design
- Engineering، Analysis، and Planning
- Tools and manufacturing.

## 2.7 Applications and Methods of Medical RP

RP applications on a range of rare medical specialties including oral and maxillofacial and dental surgery have been growing. Complex surgery in medicine requires a long period. Surgical planning can reduce the duration of surgery to reduce the incidence of risks and complications. The advances in the techniques and methods of radiography used and rapid medical prototypes and using RE can build 3D models of anatomical structures of the body to support the planning process. Visual representation of objects allows a surgical simulation of surgery before surgery. The survey data collected on the patient are taken by means of magnetic resonance imaging (MRI) technology. RP features such as the exact match of the 3D medical image data that have been designated as a physical model that the surgeon can preview and touch. 3D models use anatomical structures for preoperative planning, disease diagnosis, simulations, medical device models, and medical devices. Medical prototypes are used to plan surgery and reconstructive agriculture for facial, maxillofacial and spinal surgery [15].

## 2.8 RP techniques

Rapid prototyping or additive manufacturing describes a process by which a product derived from a CAD is built in a layer-by-layer fashion. In contrast to conventional manufacturing processes like injection molding, 3D printing has introduced an era of design freedom and enabled the rapid production of customized objects with complex geometries. One of the major advantages of 3D printing is the capacity to directly translate a concept into a product in a convenient, cost-efficient manner. It eliminates the typical intermediary stages involved in product development, such as development, production, assembly lines, delivery, and warehousing of parts, and the subsequent savings made from using fewer materials and labor lead to an overall reduction in the cost of production. 3D printing has been utilized in industrial design since the 1980s; however, it has only become adapted for medical applications in the last decade. Imaging data from routine (CBCT) or (MRI) can be converted into a CAD file using a variety of 3D software programs, such as 3DSlicer. These files are processed into data slices suitable for printing by proprietary software from the 3D printer manufacturers. While a range of 3D printing techniques has been developed for industrial use, Stereolithography (SLA), Multijet Modeling (MJM), Selective laser Sintering (SLS), binder jetting, and Fused Deposition Modeling (FDM) [16].

Table (2-1). The most commonly used 3D printing techniques in medical applications.

| RP Techniques | Pros | Cons |
|---|---|---|
| SLA | Current gold standard. High resolution Increased efficiency with an increase in print size. Detailed fabrication of internal structures | Day of printing time required. Require extensive post-production manual handling. High cost related to the materials, and the maintenance |
| FDM | Low cost. Minimal maintenance .High availability of printers | Require post-production manual removal of support structures Poor surface finish Mono-color and mono-material with the current technology |

## 2.8.1 Stereolithography (SLA)

The term Stereolithography (SLA) was first introduced in 1986 by (Charles W. Hull), who defined it as a method based on photopolymerization of liquid monomer resin for fabricating solid parts. An SLA device typically consists of a vat filled with liquid resin (acrylic or epoxy resin), a movable elevator platform inside the vat, an ultraviolet laser and a platform, as shown in Figure (2-3). In SLA, the laser beam following the path defined in the slicing model cures the surface layer of resin selectively. After creating one layer, the movable platform is lowered into the vat, and then the laser beam tracing process is repeated. This process continues layer by layer until the part fabrication is completed. The advantages of SLA include smooth surface finish and high part building accuracy, whereas disadvantages of this process include time-consuming post-processing and the use of expansive and toxic material. SLA is well suited for craniomaxillofacial surgery (congenital, system-bound growth disorders and facial craniosynostoses). This technology (SLA) also has a strong perspective for biomedical applications, such as to manufacture anatomically shaped implants, tailor-made biomedical devices and has proven to facilitate and speed-up the surgical procedures, especially in implant placements and complex surgeries [17].

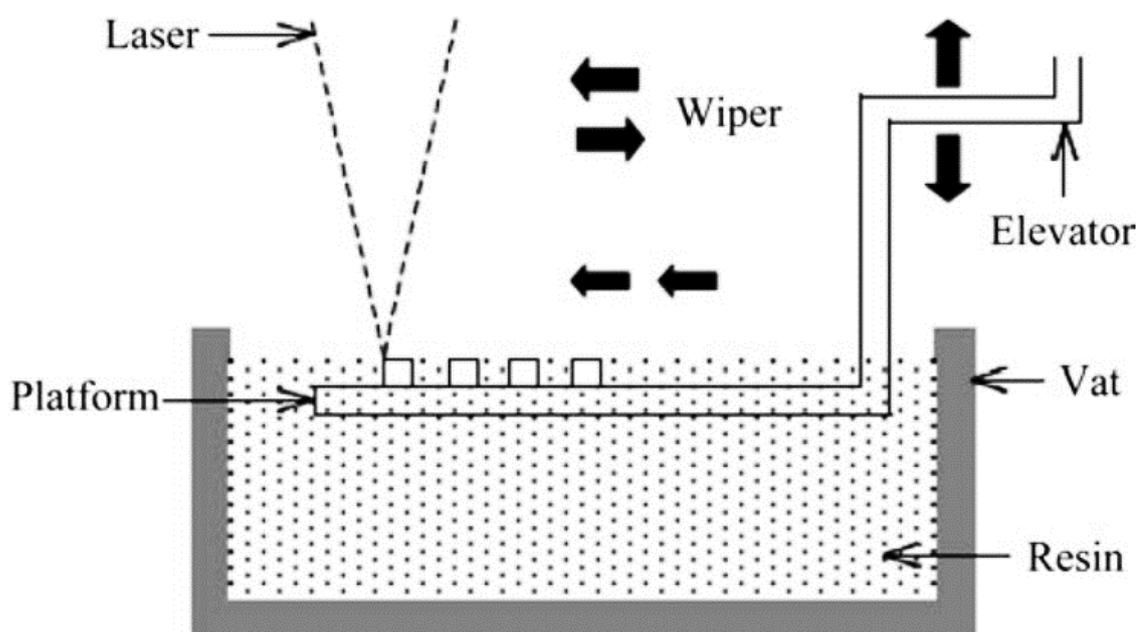

**Figure (2-3): Schematic of SLA process** [17]

## 2.8.2 Fused Deposition Modeling (FDM)

FDM technology is a method for manufacturing additives based on melting and polymer depositions more of cry, the material layers are combined together to form an object. The material is melted at a temperature of up to (0.50 ° C) or slightly higher. It is then passed through a copper nozzle in which the molten liquid is formed in the form of a continuous line that reaches a drop of less than 0.2 mm and the nozzle moves according to a path drawn by the computer for each layer separately as shown in figure (2-4). The device moves to the top layer and builds it in the same way. Using materials such as ABS, wax, chocolate, and others. FDM is the simplest way to achieve 3D printing and is widely accepted in that it is cheap and efficient [18].

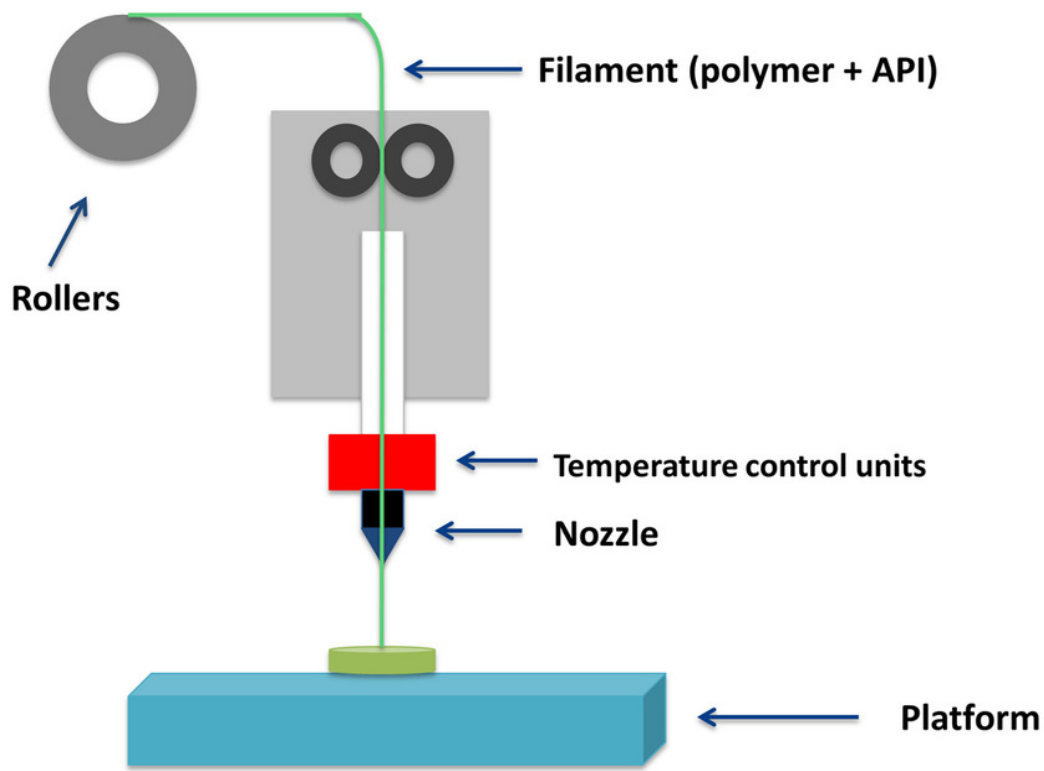

Figure (2-4): Schematic of FDM process [18].

# CHAPTER THREE

Literature Survey

# CHAPTER THREE

# Literature Survey

## 3.1. Introduction

Current research has not given a realization of the integrated environments between RE and RP technology, most of the research highlights that this relationship consists of small case studies of technology. Therefore, fabrication managers and Technicians in the medical field have limited experience of the types and methods of integration of rapid technologies and their impact when combined.

This chapter shows a review of relevant literature researches on possible technologies and applications of RE and RP in the medical context. The overview of the research topics is divided into two parts explained later on. The First part includes RE Image Processing and 3D Modeling of Mandible Reconstruction. The second part Mandible Reconstruction Design Analysis Simulation and RP.

## 3.2 RE Image Processing and 3D Modeling of Mandible Reconstruction

Many methods of modeling allow dealing with medical images and turn them into 3D models as pointed out by many researchers. This leads to a direct interaction between the geometry of the mandible and the diagnostic plan on the one hand and the model of simulation of surgical intervention, on the other.

**J. V. L. Silva et al. 2004** [19]**:** Developed a methodology by using the RP technique for selecting the biological materials needed to treat deformities and damage the bruises caused by accidents. The main objective of this work is to provide and integrate computer systems, methodologies, and the use of RP technology to reduce the cost of post-surgery and risk. Regular surgical planning and selection of vital material are appropriate to build a dedicated implant for the patient.

**John Winder and Richard Bibb 2005 [20]:** Focused on reviewing modern programs and devices for the purpose of modeling and manufacturing high-quality medical models using RP technology. The software used in the modeling and 3D modeling process helps to eliminate noise in the patient's image caused by the artifact within the jaw. Medical models based on SLA technology and FDM technology are designed for a wide range of facial, jaw, orthopedic and other technologies side by side.

**Douglas P. Sinn et al. 2006 [21]:** The development of CAD techniques has helped to build 3D models for diagnosing complex congenital malformations. It also helped to provide additional information necessary for patient planning and correction of distortions very accurately. In addition, the pre-operative planning added to the doctors' assistance in developing an integrated plan on the procedures of the process by printing a 3D model. This research highlights the evolution of technology and its use in the operation of cranial facial surgery and provides an example of its use.

**Jelena Milovanović et al. 2007 [22]:** Reviewed the procedures for fabricating medical models in different fields of medical and future trends using various RP techniques. The most specialized applications of techniques in surgical planning and treatment were used for complex operations procedures as well as training, simulation, medical instruments and diagnostics.

**Esfandyar Kouhi et al. 2008 [23]:** Presented a method for the designing and manufacture of special medical forms using CAD and RP techniques for application in reconstructive jaw surgery. The proposed methodology was to take data from computed tomography (CT) and data processing using a CAD model and then to examine the effect of advanced processing in RP methodology.

**Sekou Singare et al. 2009** [24]**:** Described preoperative planning and computer-assisted design based on the patient's image data using the CT scan. During the design operation, the correct area of the bone part determined based on the visualization technique and was manufactured using RP techniques.

**Ihab El-Katatny et al. 2010** [25]**:** Presented a review of the errors resulting from the fabrication of complex anatomical models based on computerized tomography (CT) and FDM. Eleven parameters compared to virtual models and eight parameters selected to evaluate bone thickness variation. The results showed a standard deviation of 0.11% for the mandible and 0.16% for the skull. In addition, high accuracy obtained for the manufacture of anatomical models using FDM.

**Li-bin Zhou et al. 2010** [26]**:** The use of reverse engineering, CAD models, and the manufacture of jaw treatment using techniques RP is being studied for improving the surgical operation of the mandible with high accuracy. The reverse side technique used for patching the damaged side with a replica.

**Eero Huotilainen et al. 2013** [27]**:** Based on computed tomography and patient medical data, three-dimensional physical models of the skull have been created. The aim of the study was to apply the RP technique to demonstrate the differences resulting from the conversion of DICOM by dedicating software to STL in the medical model.

**Emad Abouel Nasr et al. 2014** [28]**:** Progress in medical imaging, computer-aided design, and RP techniques have helped to design and manufacture models for anatomical structures and patient implant fabrication. The results of a continuous project in jaw reconstruction are still using the technique of manufacturing electron-fusing (EBM) additives.

**Maria Aparecida Larosa et al. 2014** [29]**:** Employed an Additive manufacturing technology to help in the production of prostheses of different types and sizes as well as obtaining prototypes that contribute to surgical planning.

**Jardini, A.L. et al. 2015 [30]:** Designed and manufacture biomedical implants for a patient with a major skull defect. The modeling process based on the patient's medical data and computerized tomography and the use of CAD programs. Using a methodology to create a biological model of bones, as well as pre-operative planning, helped to reduce the duration of surgery and improve surgical accuracy.

**Maureen van Eijnatten et al. 2016 [31]:** Studied the effects of manual threshold technique selection on the accuracy of medical model construction and conversion to STL based on the accuracy of computerized tomography. That research explains that the manual threshold technique produces more suitable STL models than the automatic threshold.

**Aditya Mohan Alwala et al. 2016 [32]:** AM technology is widely being used in maxillofacial surgery for hassle-free planning, patient education and execution of the surgical procedure and for precision using medical models. The Current case is of pan facial trauma with multiple facial bones fracture treated by surgical planning on AM medical model to adapt the mini plates to be prior to the surgery.

**Jan Egger et al. 2017 [33]:** Illustrated the uses of the (MeVisLab) platform prototype, which can be used as a successful alternative to a complex trading platform. The fictitious prototype helps the complete workflow to build the implant. Thereafter, the manufacturer can make the necessary modifications to the model for the purpose of the patient implantation.

**Zhifan Qin et al. 2018 [34]:** The 3D model can be constructed by using a "mirrored" technique of the corresponding unaffected mandible. Depending on the patient's medical data and using computerized tomography and using a 3D platform, the damaged area was isolated.

**Trajanovic, M. and Tufegdzic, M. 2018 [35]:** here is a need to examine the possibilities for AM application in the field of orthopedics and prosthetics and to provide an overview of the current state. Progress done in the area of biomaterials is presented from the perspective of different AM technologies used. Patient-specific implants and prostheses in different branches of surgery and implantology, as well as some examples in tissue engineering, are shown.

**Tahseen F. Abbas et al. 2019 [36]:** Develop a concept of 3D free form surface reconstruction rig and generate the dimension of the surface depending on the proposed image processing technique. Through the results obtained are observed that. The overall calibration accuracy of the root means square (RMS) value.

**Maorui Zhang et al. 2019 [37]:** Based on a clear methodology that begins with the acquisition of the radiograph of the patient and the formation of models for planning surgery and treatment. The results were satisfactory after a follow-up of patients who formed the iliac flap to reconstruct the mandible defects. The study showed an ideal bone union mandible.

## 3.3. Mandible Reconstruction Design Analysis Simulation and RP

Recently, the application of CAD system and RP techniques in addition to FEA in the reconstruction of the mandible has drawn the attention of many researchers as show later on and it became necessary to apply these techniques in the description of treatment for patients.

**I. Gibson et al. 2006 [38]:** Describe cases in which RP technology is used in medical applications. This is done by addressing a number of medical studies that demonstrate the overall uses of the RP technique. Accordingly, these studies analyzed and applied in the medical field for describing the treatment necessary to solve the related medical problems. As a result, the study showed that RP was very useful in solving medical problems despite many limitations.

**Jiman Han and Yi Jia 2008 [39]:** The 3D models of body structures from CT images are reconstructed in SolidWorks by using the (Planar Contour) method and Medical Rapid Prototyping (MRP) models are performed.

**Wei Zhong Li et al. 2009 [40]:** Used computerized CT scans, the deformities experienced by the patient with multiple fractures are reconstructed in the ZOMC area. The CAD system helped to design the distortion area using Mimics 10.0. The skull model has been reproduced using RP technology.

**Liliana Beldie et al. 2010 [41]:** Based on the patient's medical data, the medical model created. The model was then used to simulate and plan the surgical and simulation of facial changes before and after surgery. The results were astounding, with an 85% overall agreement with patient data. Based on these results, the proposed methodology and planning tools can apply to establish a detailed and accurate medical model.

**P. U. Ilavarasi and M. Anburajan 2011 [42]:** Used a CAD system, a 3D model analyzed and constructed. Using ANSYS software, the mechanical properties and suitability of the implant are examined. The results were as expected as the measured displacement and maximum value became -16.1% and -4.8%, respectively. The mechanical behavior of pressures proved that the reconstructed mandible became close to the normal bone.

**Kaufui V. Wong and Aldo Hernandez 2012 [43]:** Presented model within the CAD system and then converting the file to STL for 3D printing. The surfaces rounded by triangles and the model cut into slices containing the information of each layer being print. This technique has made medical practice and studying designs by engineers very easy.

**Hind Basil Ali et al. 2013 [44]:** Study on the influence of three FDM process parameters (layer thickness, infill density, and printing orientation) on the mechanical properties (tensile strength, bending strength and compression strength) of the FDM manufactured parts using the Taguchi method.

**Mingyi Wang et al. 2013 [45]:** 3D models were constructed based on CT data. Simulation of defect models and loading of vertical and lateral stresses of 150 N. The distribution of stress was very good and rational with the help of implantation, which bears part of the pressure on the affected side satisfactorily and is suitable for maxillary reconstruction.

**Konstantinos Salonitis et al. 2015 [46]:** Modeling methodology for a process chain consisting of laser cladding, as an AM process, and high-speed machining (HSM), a subtractive process usually used for finishing operations. The modeling methodology, which is based on the finite element method and utilizes the level set method to define the cutting tool path, is able to predict results such as residual stresses and part distortion. The proposed approach is applied in simulating HSM of a steel tube fabricated by laser cladding.

**Sunpreet Singh et al. 2016 [47]:** The most widely used AM techniques for biomedical applications. Special attention has been paid on Fused deposition modeling (FDM) based AM technique as it is economical, environmentally friendly and adaptable to flexible filament material. This review paper will be helpful to the researchers, scientists, manufacturers.

**Mazher I. Mohammed et al. 2017 [48]:** Examined an algorithm to design the implantation of each patient based on medical imaging data. For left side design, right-side data are used using the reflection technique. The final medical model manufactured from medical titanium. The last transplant was extremely powerful and was ideally suited to a representative model of skeletal anatomy.

**Santosh Kumar Malyala et al. 2017 [49]:** Based on the patient's medical data, the appropriate implantation designed and manufactured as well as fixation of the missing upper teeth. The integration of techniques helped to strengthen, improve and achieve transplantation, which helped to reduce blood loss and shorten actual surgery time.

**Joel C. Davies et al. 2018 [50]:** Virtual simulation and 3D mobility templates have been used for preoperative planning. The prototype of the damaged mandible and healthy mandible bone built on CBCT imaging. The proposed techniques include digital tumor removal using a digital 3D model and the missing part of the tumor generated using a "mirrored" technique.

**A. Manmadhachary et al. 2019 [51]:** Discussed the importance of AM technology and what has been achieved in complex surgical operations to achieve better results. In order to address the deficiencies in the traditional procedures of the processes were used rapid prototypes. AM used in this research to overcome the problems and challenges of surgical operations and to educate patients.

## 3.4 Summary

The topic of reconstruction of the mandible is of great interest to researchers because of its great importance to the patient and the restoration of functions and the improvement of the external appearance. There are some points that must be concluded, as follows:

1- From the previous literature, various methodologies concerned with surgical planning and implant model design have been studied and developed.

2- This work presents a proposed methodology for reconstructing the mandible using a virtual computing environment. From this work, it understands that the preoperative planning process it is of great importance among surgeons.

3- Previous research has shown several methodologies in this field and has been compared to this work. A virtual computer environment has not been used, while the effect of this environment on this work will be studied and its effect studied. The implant will be designed for the patient in a different way for many researchers where the mesh design is used.

4- It can conclude based on the literature review that there is still enough scope for research in order to improve surgical methods and preoperative planning by improving the methodologies used in constructing the implant in this work.

# CHAPTER FOUR

# The Proposed Methodology

# CHAPTER FOUR
## The Proposed Methodology

## 4.1-Introduction

This chapter describes a proposed methodology for developing an integrated environment between RE and RP. For the purpose of reconstructing the mandible and doing simulation and surgical planning. Traditional methodologies are used to design and manufacture a medical model based on patient medical data. Surgical simulation performed on a 3D physical model made with RP techniques. To improve the level of simulation and surgical planning, engineering programs are used for image processing, modeling, and medical simulation. The combined integration of engineering software will significantly reduce the time of surgical intervention. Surgery in a virtual computer environment that simulates clinical reality will contribute to reduce surgical risks and improve the level of surgery. Figure (4-1) shows the main stage of applying the methodology.

**The stages of the proposed methodology as following:**

1-Acquisition of patient data

2-Image processing.

3-Customized implant design and FEA.

4- Virtual surgery.

5- RP medical model.

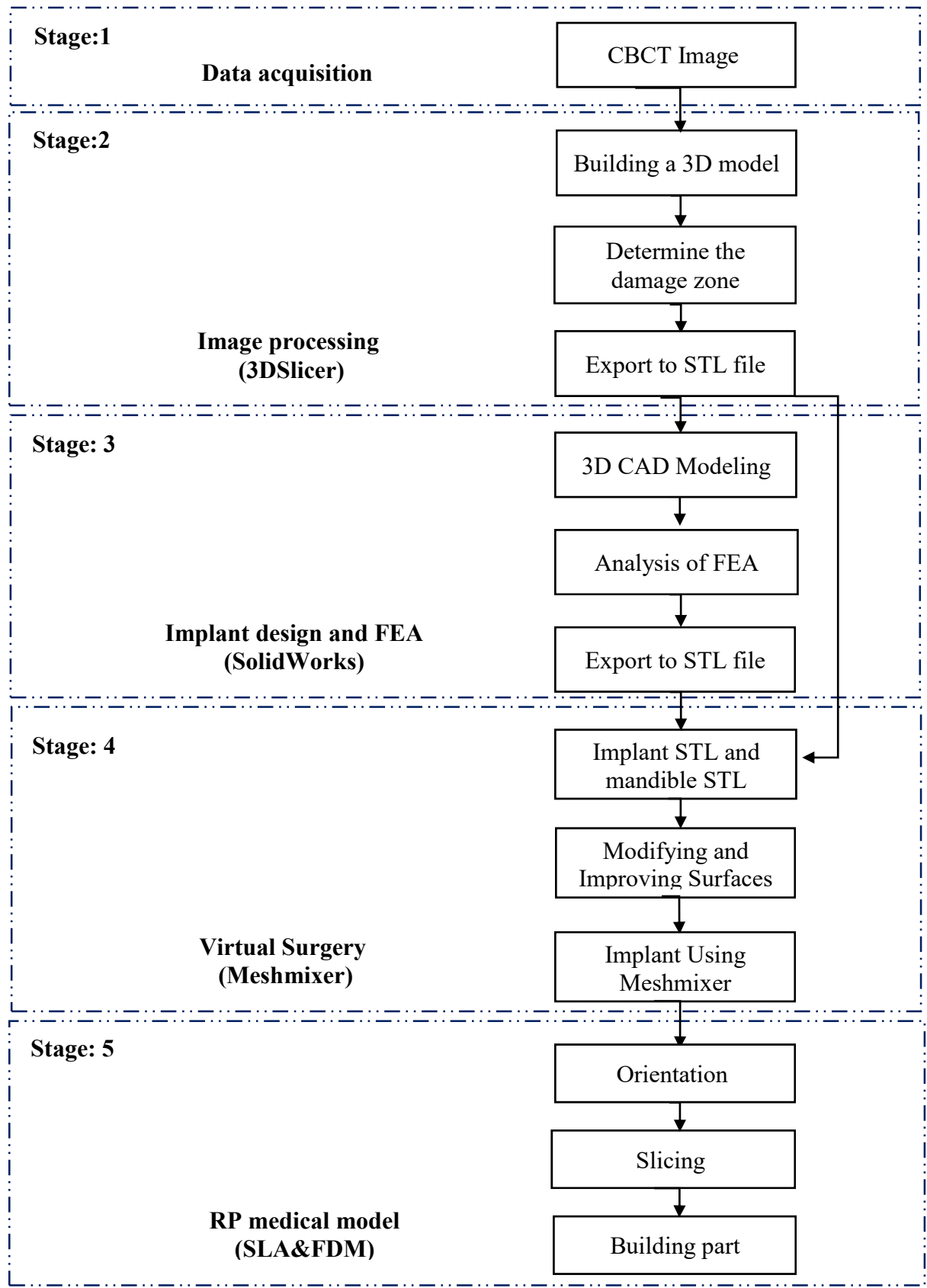

**Figure (4-1): The block diagram of proposed methodology**

## 4.2. Image Capturing

In the first stage, the patient's medical data are collected using the most commonly used techniques for capturing medical data. Which provides human body scans, namely MRI, CT, and CBCT technique. In this work, a more sophisticated type of scanner with low radiation ratio, CBCT is used. This technology allows faster scanning than conventional scanners as well as smaller scanning areas as shown in Figure (4-2). The accuracy of creating a 3D model depends on the accuracy of the scanner and the data it collects.

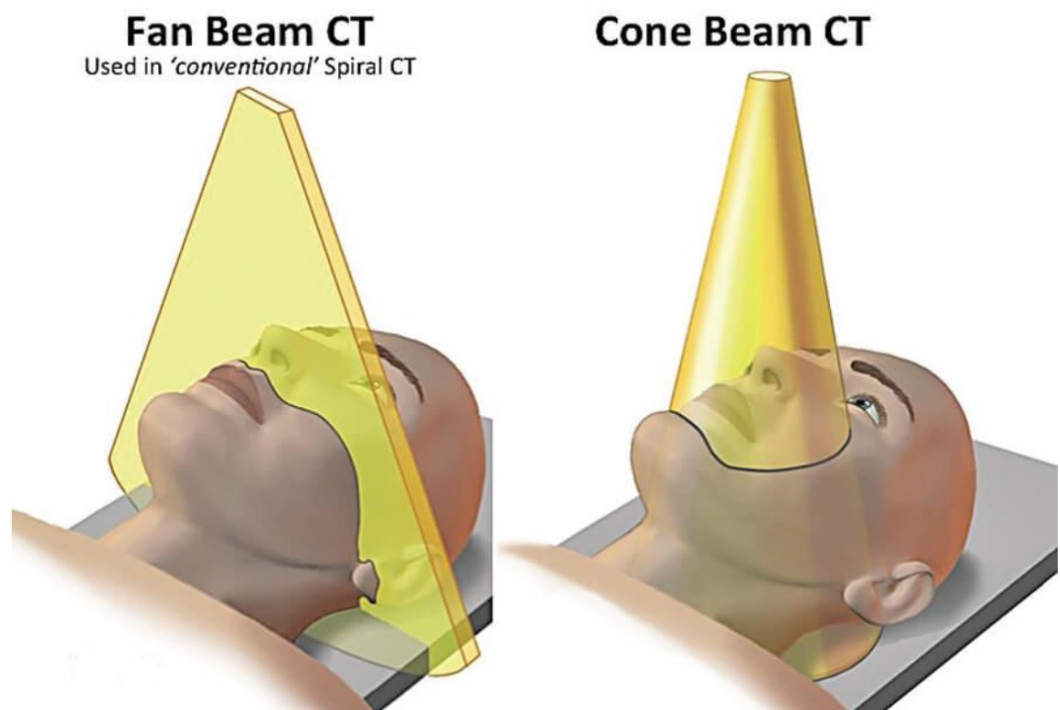

**Figure (4-2): Diagrammatic representation of image capture technique of CT and CBCT device** [52]

## 4.3. Image Processing

This stage is divided into three steps, first selecting the construction method of the model as well as identifying the damage area and methods used and finally exporting the model as an STL file will be discussed as follows.

## 4.3.1 Building a 3D Model

Based on the patient's data, obtained by a CBCT scanner, the data contain the skull and the affected area. 3D models are used in the construction of various techniques including segmentation, region growing, automatic threshold, and manual threshold. In this work, the manual threshold technique is used as shown in Figure (4-3). Because of its importance in isolating the bone area. The image must be very small, contiguous and overlapping, which is achieved by the manual threshold. Other features include imaging units and Hounsfield units that range, for example, between (-1000 and + 1000) according to tissue absorption of radiation, give bone density and bone quality. Each normal tissue in the body has its own CBCT number or Hounsfield unit that distinguishes it from other tissues, and any abnormalities in this figure indicate a defect in the area.

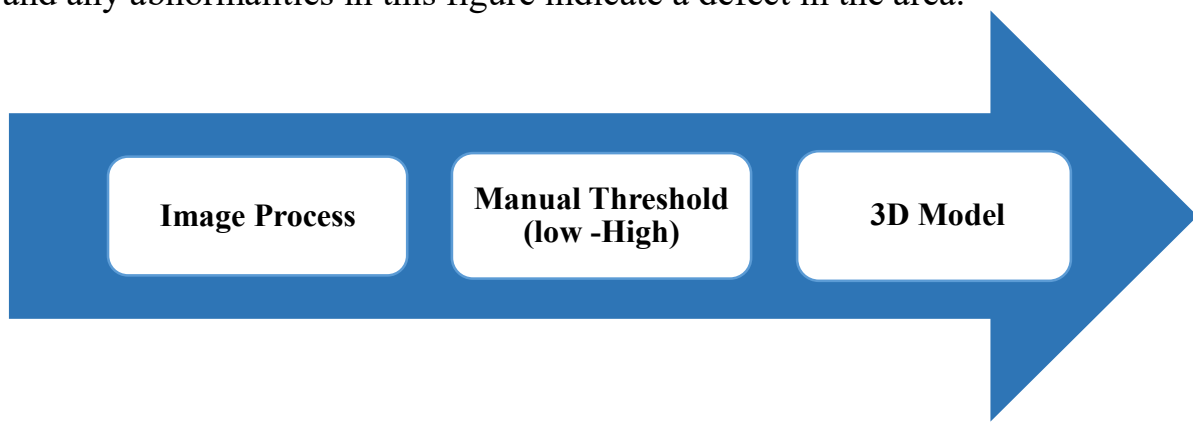

**Figure (4-3): The manual threshold method**

## 4.3.2 Determine the Damage Zone

There are several methods and techniques to identify the affected area and remove unwanted tissue. One of these methods is to determine the area of damage by the CBCT device, where the device is set approximately in the damaged area. In addition, the area determined before the fragmentation is performed using the automatic threshold. The model is then constructed and the unwanted tissue cut and wipe by a clean mask. In this work, dead and scattered bones around the area are manually removed using scanning and cutting techniques. Sometimes, 2D image segment data from scans contain previously implanted metal parts. Causing severe flashes and noise in the area around the damage, resulting in unnecessary protrusions of up to (1 cm). This condition is handled by a CBCT scanner or by using a reconstruction program.

### 4.3.3 Convert to STL File

After finishing the modeling process and building the 3D model, the model is exported in STL format as shown in Figure (4-4). The STL has several acronyms "Standard Triangle Language and Standard Mosaic Language". This file format helps many other software packages; widely used in RP, 3DP, and CAM. This coordination has several advantages, including the approximation of the stereoscopic design and simplified by dividing the surfaces into triangular areas that ultimately produce a full stereoscopic. Most 3D design programs can also produce this type of file that helps to print the model.

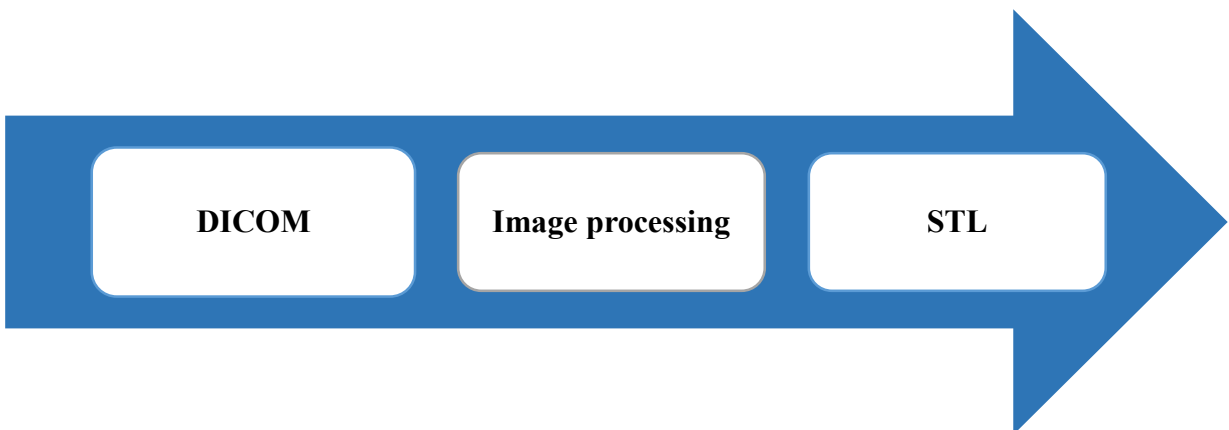

Figure (4-4): Convert to STL file

### 4.4. Customized Implant Design and FEA

At this stage, the implant designed for the affected area. Carrying out loads and implant tolerance. This is done by using the engineering software package. This phase is divided into three steps as described.

### 4.4.1 A 3D CAD Modeling

Modern technology in medicine has helped to design and manufacture patient transplants. The model building relies on reverse engineering techniques to obtain patient medical data collected from scanning techniques. There are many specialized programs in the field of design and analysis in addition to many methods of drawing and design. In this work, the SolidWorks program has been selected and developed by (SolidWorks Corp2015). This program has very high 3D design possibilities as well as the FEA analysis process, which is one of the most important features in the design process of an implant.

## 4.4.2 Analysis of FEA

"Finite Element Analysis (FEA) is a computer-based digital technique for calculating the strength and behavior of structures under different stress conditions". The FEA technique as shown in Figure (4-5), shows the power distribution associated with the parts to be reconstructed such as the skull, mandible, and maxilla. This procedure is performed to verify the strength of the mechanical properties of the implant resulting from the strength of the mastication reaction, pressure, movement, and other stresses. There are many software packages for item analysis such as ANSYS, MIMICS, and SolidWorks. In this work, SolidWorks modeling and simulation program selected for the apparent application of surgical compensation technique and mechanical stress analysis of the implant model.

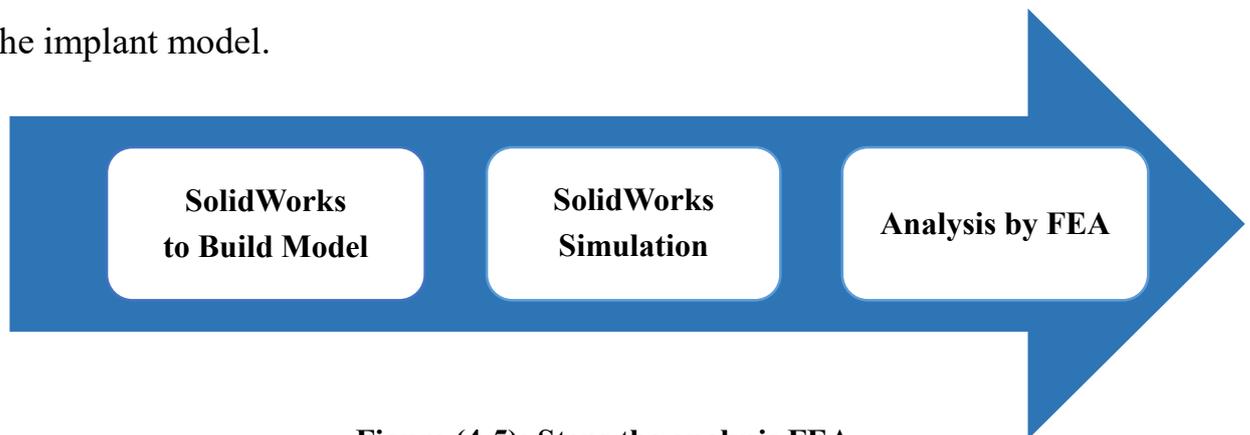

Figure (4-5): Steps the analysis FEA

## 4.4.3 Convert to STL File

Finally, after testing and analysis, the solid 3D model exported to STL format. Data should be transferred in the form of agreed formats such as STL or IGES. Defects of models stored in STL format should be checked. This is due to errors in the CAD model and Meshmixer that can be used to handle errors.

## 4.5 Simulation

At this stage, a 3D model is built based on the patient's medical data combined. Showing the design of the implant that is built using reverse engineering. Appropriate modifications are also made and the design implant is performed in a virtual computing environment as described in the following steps.

### 4.5.1 Implant STL and Mandibular Model STL

Figure (4-6) shown the ability to invest engineering software for image processing, medical modeling and simulation to perform surgery in a virtual computer environment that simulates clinical reality is crucial. The advantages of this are the ability to achieve accurate industrial models with dimensions close to the loss zone. It is also possible to evaluate the technique of surgical fixation, which was not possible in conventional methods. There are several programs that act as the default environment chosen in this Meshmixer package work.

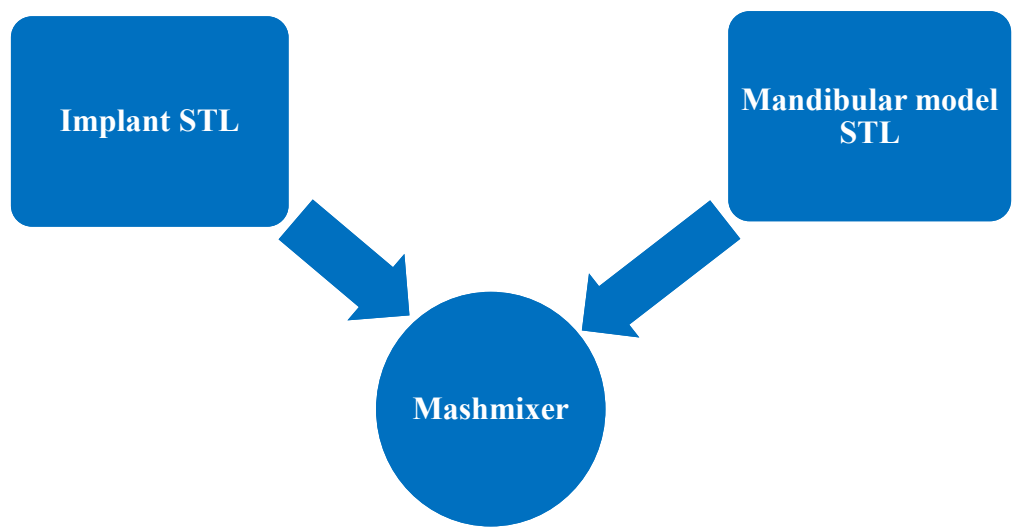

**Figure (4-6): Assemble the models**

## 4.5.2 Modifying and Improving Surfaces

The virtual computing environment proposed in this reconstruction thesis provides an overview of maxillofacial tissue. Virtual surgery enables surgeons to plan and direct surgery correctly and accurately. In this work, some necessary modifications were made to the models including reducing the size of the implant by using the cutting technique and making the surfaces softer.

## 4.5.3 Implant Using a CAD System

For the purpose of implantation of the model using the CAD environment, which is identified in the previous stage. In this work, virtual surgery and surgical planning performed prior to surgery and appropriate modifications are done as to the first step before implantation. Without the need for printing, the physical model is made of polymer and simulated. Using the (Transform) technique, the fractured part rotates at a certain angle to match the mandible to the fractured part. After implantation using (Soft Transform) technique, the implant can be easily matched with the medical model of the mandible.

## 4.6 RP Medical Model Production

This step involves selecting the appropriate RP technology to manufacture the model according to the desired purpose, as well as the required accuracy, visual appearance, and materials. In this work, SLA technology is selected for prototype printing. The 3D model must be exported in STL format to RP to create a 3D model for the patient. The quality of a typical STL file, as well as the process of routing the model within the RP device, affects the quality of the model and by specifying, the correct parameters are required to create the model.

# CHAPTER FIVE

# Work Implantation & Results Discussion

# CHAPTER FIVE
# Work of Implantation

## 5.1. Introduction

This chapter explains the application of the proposed methodology by providing an overview of the field of the case study. The case study illustrates the application of the proposed methodology through RE and RP techniques and used it in the medical field based on 2D medical images. The study included the application of integration of a design environment, analysis and rapid prototyping in the creation of new design in the medical field. The Stereolithography technique is used to manufacture the mandible model and the treatment model to clarify the proposed methodology and make the models accurate. In this work, the files stored in the smart computer as an STL file imported into Meshmixer. A program allows editing and development on 3D files and works on any operating system developed by Autodesk. Finally, the results are discussed.

## 5.2. Overview of the Case Study

According to the agreements between universities and medical cities (teaching hospitals), the selected case study is the Medical City in Baghdad. In Martyr Ghazi Hariri Hospital / Oral and Maxillofacial Surgery Department / Radiology Unit and 3D Printer, Iraq. A survey CBCT was conducted to assess the integrity of the skull and other tissues of the patient. The research sample selects a 35-year-old male, who suffers from a shocking car accident and this is resulted in significant in the anterior teeth of the maxilla and the loss of mandible large part with the teeth.

## 5.3 Work Implementation

The work implementation of the case study was being carried out through five phases as shown in Figure (5-1). These phases started with collecting medical data of the patient and finished with a presentation of the prototype.

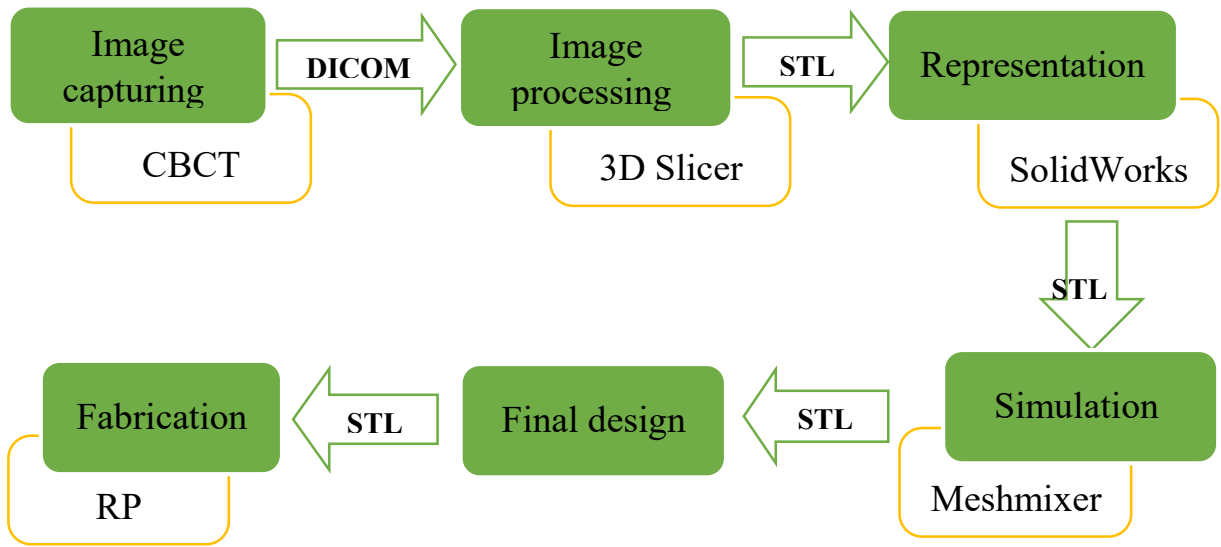

Figure (5-1): The stages of implementation the work

## 5.4 Image Capturing

The CBCT technique helps to obtain smaller and faster scan distances compared to conventional scanners as shown in Figure (5-2). Parameters of the spiral scanner are set to peak voltage up to (120 kV). In this work, the voltage (70 mA), the spacing (voxel gap) is (x, y, z) respectively (0.32, 0.32, 0.32) the dimensions of the image (x, y, z) respectively (468,468,407), and the intensity range (- 1000 to 3095) this image is saved DICOM.as shown the figure (5-2).

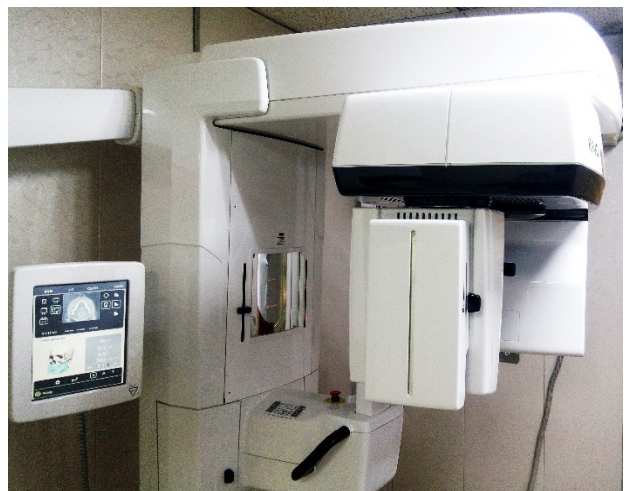

Figure (5-2): CBCT scan OnDemand3D

## 5.5 Image Processing

Within the proposed methodology program has been chosen (3DSlicer). Data is entered from the (DATA) command in case the device not connected to the computer and the (DICOM) command in case the device is connected to the computer. The (VolumeRendering) menu contains commands that are required for interacting with controllers. Currently, choose the (Volume) command to display the 2D image in three levels top, side and back as shown in Figure (5-3). The (Preset) command, which contains a set of predefined functions that maintain opacity and color gamut. Besides, select the (Shift) command to determine the level of bone tissue density induced by manual selection.

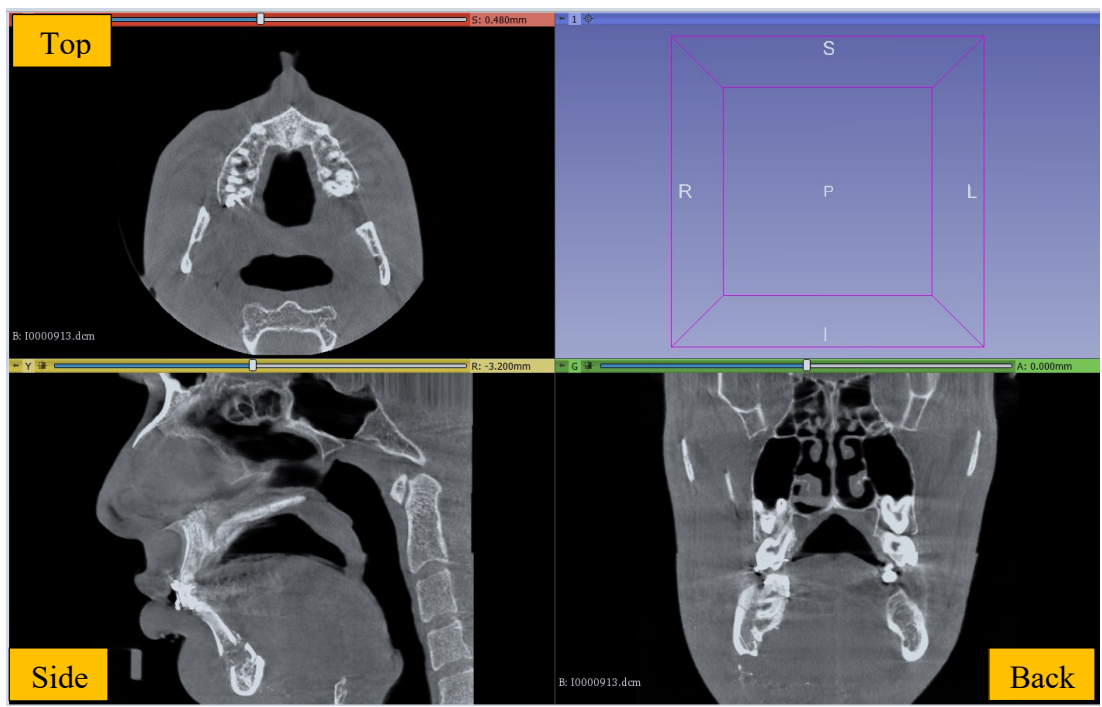

**Figure (5-3): 2D images distributed over the three levels top side and back.**

Figure (5-4) shows the (Crop) command and simple controls for the crop box (ROI). Which allows cropping unnecessary excess tissue. Currently, the command (Crop Volume) has to be select from the list of (ALL Modules) and apply to the modifications performed to the prototype to reduce the overall size of the skull. As shown in Figure (5-5).

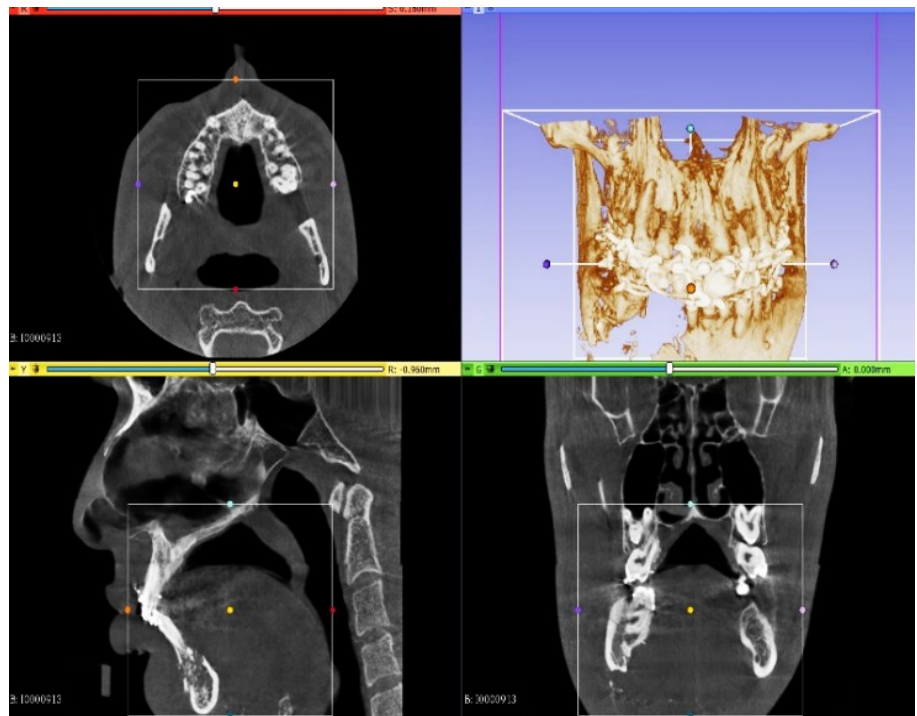

**Figure (5-4): Illustrates a command (Display ROI)**

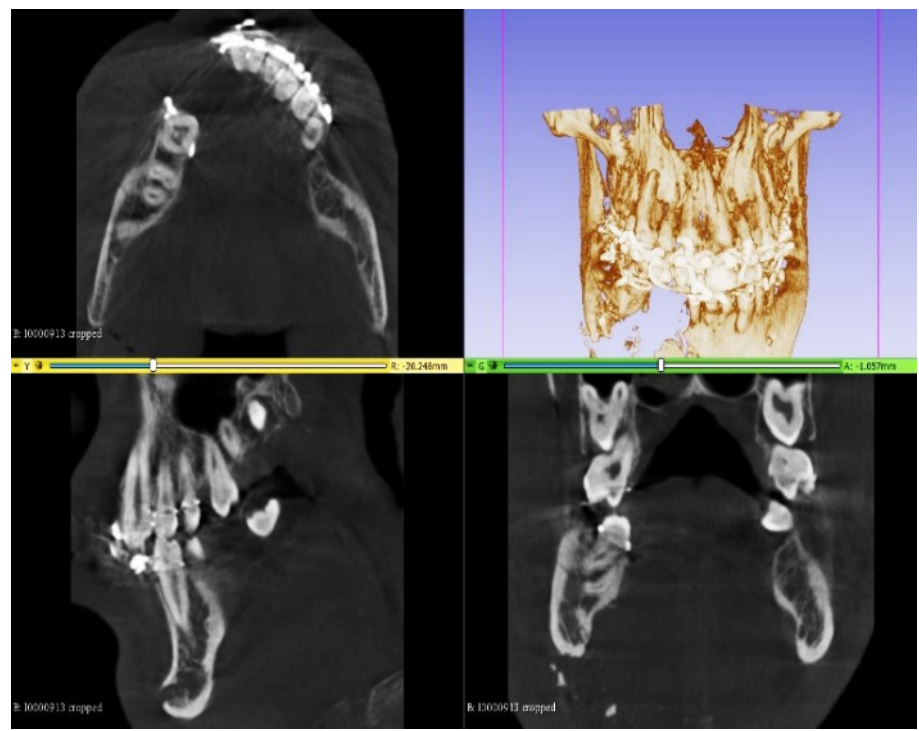

**Figure (5-5): Shows the selection of (Crop Volume)**

Figure (5-6) shows the (Slide Editor) command selected from the menu (All Modules). The form will be added to (Segment Visibility in 3D) from (Add). In this process, the desired fabric color is determined. In this case, choose the yellow

color that represents the bones. Figure (5-7) illustrates the use of a manual threshold technique in building the 3D model, which ranges from (355.44-2677.32) HU unit value. Furthermore, the removal of fragments is scattered by injury using (Erse) to clean the model.

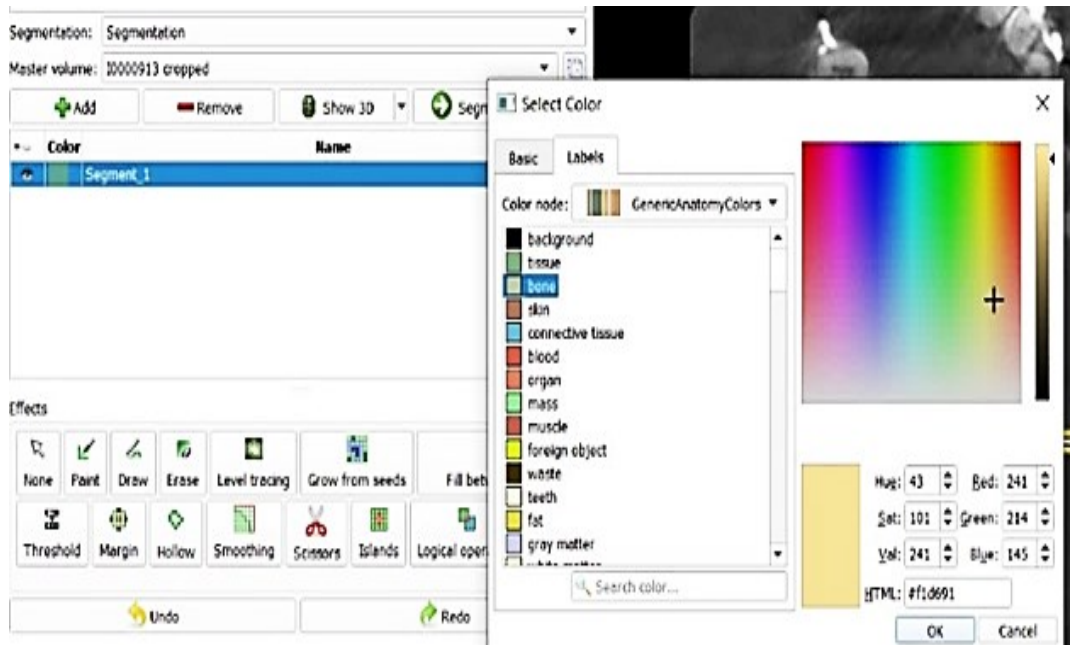

**Figure (5-6): Shows the selection of tissue type and color.**

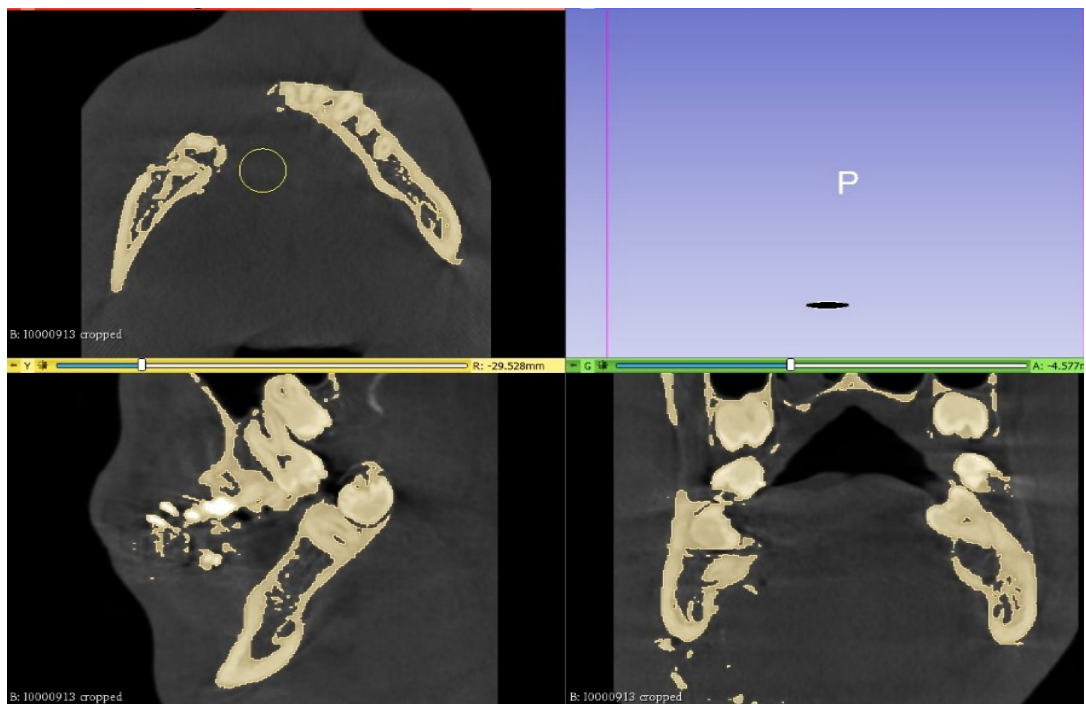

**Figure (5-7): Illustrates the threshold technique.**

Figure (5-8) illustrates the process of building a 3D model using a manual threshold technique. Finally, expand the export/import section at the bottom of the hash unit select the output and then export. Saves in the STL file format.

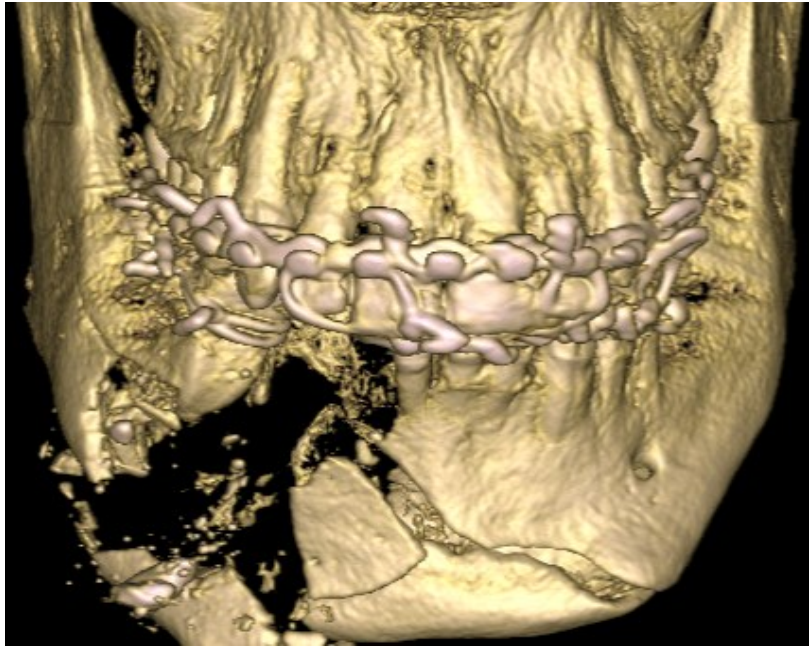

**Figure (5-8):  Building a 3D model**

## 5.6 Customized Implant Design and FEA

This stage consists of two first paragraphs, designing the missing part using Solidworks and the second paragraph of Finite Element Analysis.

## 5.6.1 Three Dimension CAD Design

Initially, the SolidWorks program opened and selected the window (New SolidWorks document), and select (Part). Then, choose the plane to select the (Front Plane). After that, the drawing on the image is selected from the menu (Tools) and, (Sketch Tools) and then chooses the (Sketch Picture). The model imported into (Meshmixer). This software allows for obtaining a 2D image of the loss zone. The researcher drew the area of loss with approximate dimensions and is not a fact using the technique (B-spline) drawing on the image of the model shows Figure (5-9).

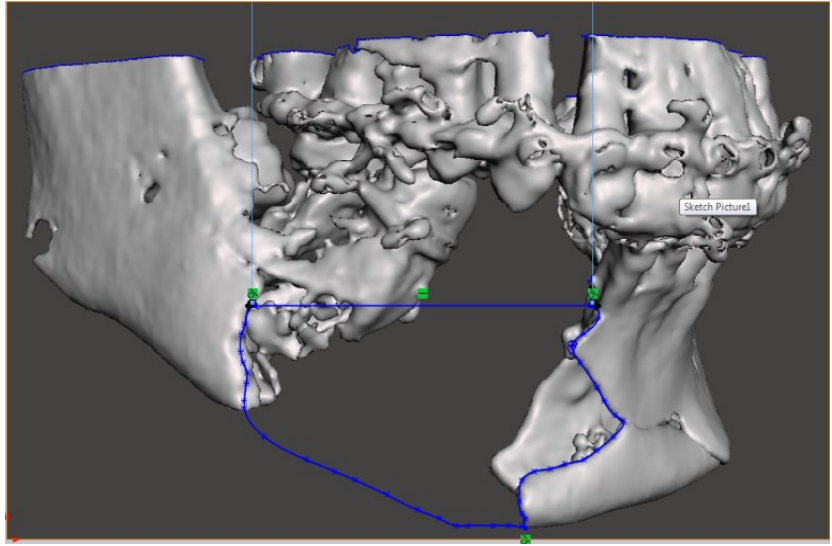
**Figure (5-9): Drawing on the image**

To convert the design to 3D, use the (Extruded Boss/Base) command and specify the required thickness as shown in Figure (5-10).

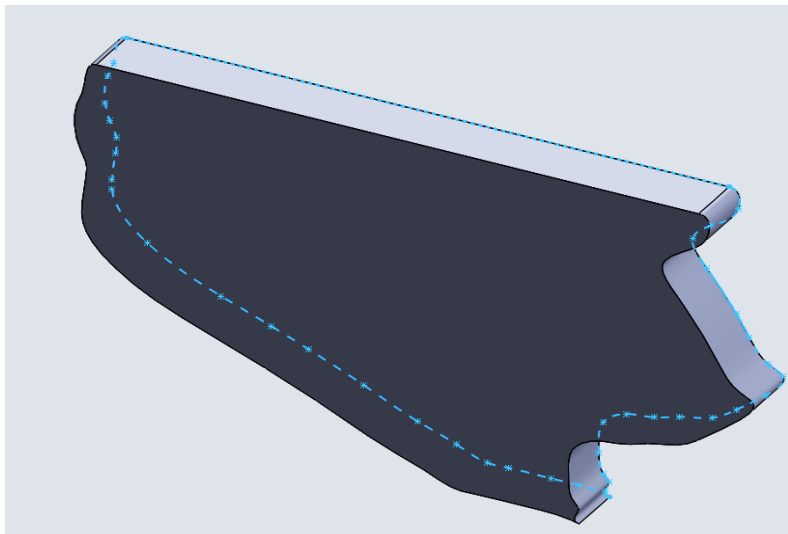
**Figure (5-10): Extruded the model design**

Now, use the command (Flex) as shown in the figure (5-11). From a list (Features) to rotate the design at an angle of 84° for making a rough approximation of the angle of the mandibular deviation .

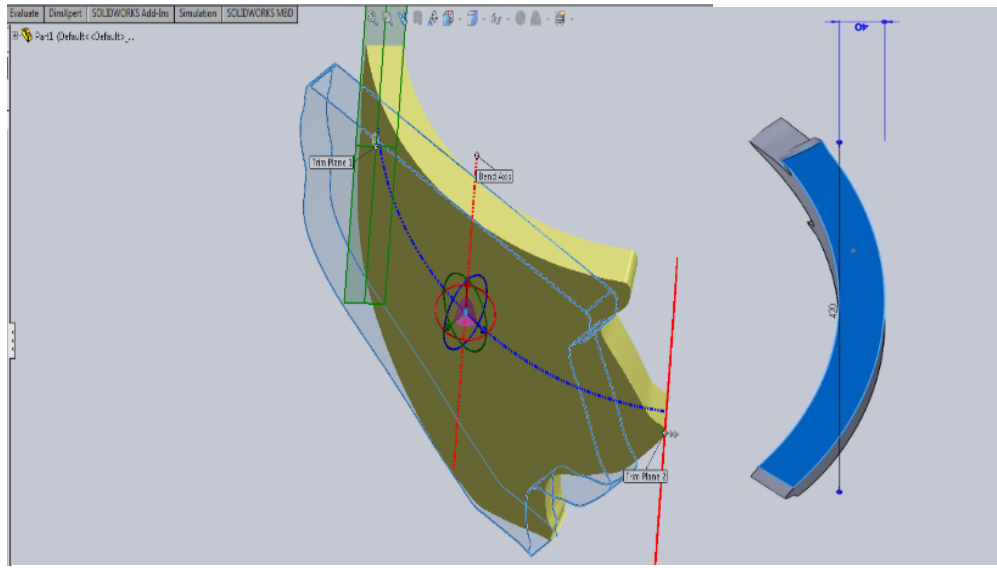

**Figure (5-11): Used the command (Flex)**

## 5.6.2 Finite Element Analysis

For verifying the durability of the final design and feasibility by using FEA to check the optimal distribution of force that is likely to shed on the model after implantation. This process is very important for making sure that there is no weakness in the design, structure, which can lead to major failures. Therefore SOLIDWORKS (simulation) is used to analyze the elements of the design model. According to the final report of the analysis, the model information recorded as in Table (5-1).

**Table (5-1): The model and properties**

| Document Name and Reference | Treated As | Volumetric Properties | Document Path/Date Modified |
|---|---|---|---|
| Flex1 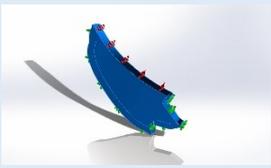 | Solid Body | Mass:0.199127 kg<br>Volume:0.00449621 m^3<br>Density:4428.78 kg/m^3<br>Weight:195.145 N | D:\Samples\FEA.SLDP<br>May 28 14:59:12 2019 |

Then, the properties of the studied material were determined, and the appropriate loads and fixing areas were applied to the design model as shown in the following table.

**Table (5-2): Material Properties**

| Model Reference | Properties | Components |
|---|---|---|
| 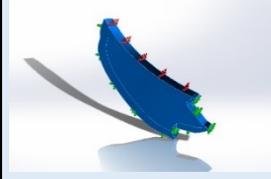 | Name: Ti-6Al-4VSolution treated and aged (SS)<br>Model type: Linear Elastic Isotropic<br>Default failure criterion: Max von Mises Stress<br>Yield strength: 8.27371e+008 N/m^2<br>Tensile strength: 1.05e+009 N/m^2<br>Elastic modulus: 1.048e+011 N/m^2<br>Poisson's ratio: 0.31<br>Mass density: 4428.78 kg/m^3<br>Shear modulus: 4.10238e+010 N/m^2<br>Thermal expansion coefficient: 9e-006 /Kelvin | Solid Body 1(Flex1)(FEA) |

This work is considered very important because of the force that sheds on the mandibles of the maxilla such as the force of chewing and cutting. On this basis, a fixed force with similar limitations to the chewing force projected at the actual implant position see Figure (5-12A). Figure (5-12B) shows the application of the mesh to the implant after the installation of load restrictions.

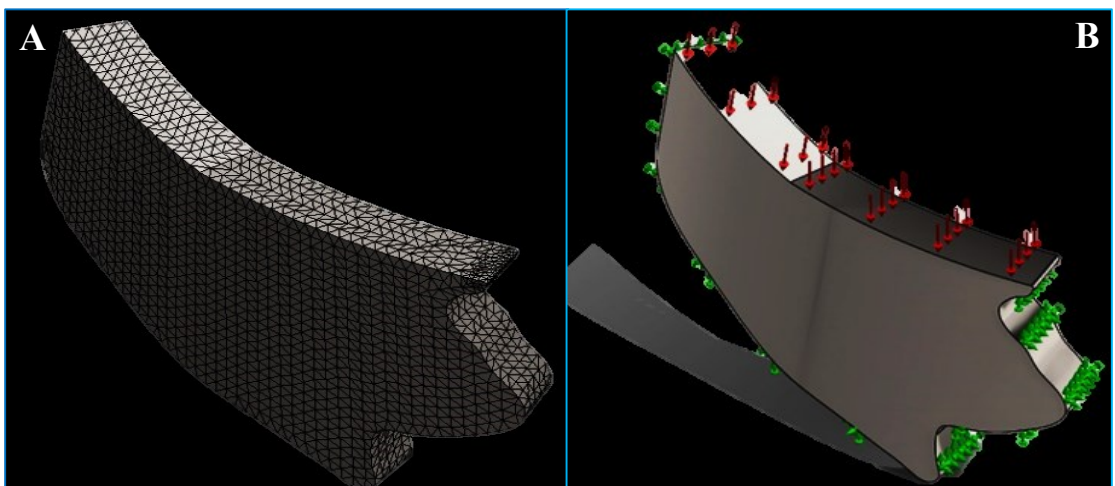

**Figure (5-12): The FEA (A) a mesh model of the implant (B) the constraints and loads**

Figure (5-13) shows the analysis of the distribution of stress values on the design because of applying a force of up to 200 MPa.

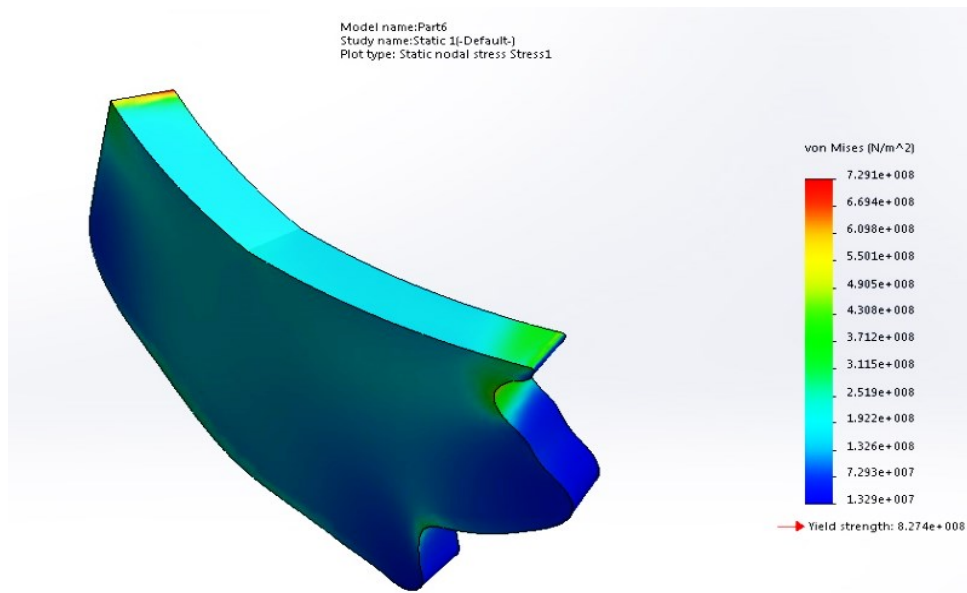

**Figure (5-13): The FEA Color-coded of the implant**

Table (5-3) shows the results of the maximum values of Von Mises stress, displacements and the maximum strain of the design. Stress values are included in the design according to the expected force on the model to study the effect of pressures. Von Mises yield can be expressed by the following equation.

$$(σ1 − σ2)2 + (σ2 − σ3)2 + (σ3 − σ1)2 = 2(σy)2$$

**Table (5-3): loads and fixtures**

|  | Loads (MPa) | Max von Mises Stress N/m^2 | Vector sum of displacement (mm) | The Max value of strain |
|---|---|---|---|---|
| The mandible reconstruction | 100 | 3.64233e+008 | 0.221893 | -0.221893 |
|  | 200 | 7.29051e+008 | 0.443708 | -0.00510231 |

Finally, after the completion of the analysis process, the model is exported as a stereoscopic file (STL). This is one of the most common formats used in 3D design. STL Reconstruction is then, imported into Meshmixer where it can be finalized before printing.

## 5.7 Virtual Surgery and Implantation

Figure (5-14) illustrates the process of importing a 3D STL file created based on patient data and a 3D implantation STL file designed in a CAD environment based on the missing portion of the patient's mandible. Ripples and overlapping surfaces observed due to the different design environment as shown in Figure (5-14). Accordingly, from the (View) menu, a selection of (Show Wireframe) to show the size of the triangles as shown in Figure (5-15 A). The surfaces are treated and the triangles reduced to give smoothness and bonding to the surface using the (Reduce) button from (Sculpt) menu see Figure (5-15 B).

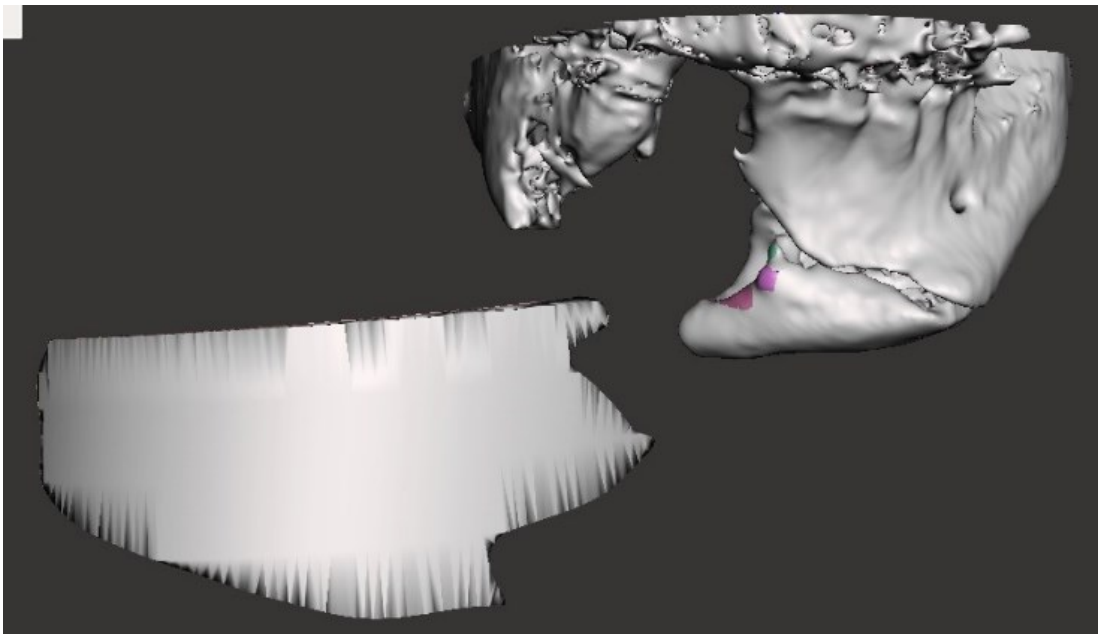

**Figure (5-14): STL file for mandible model and STL file for design**

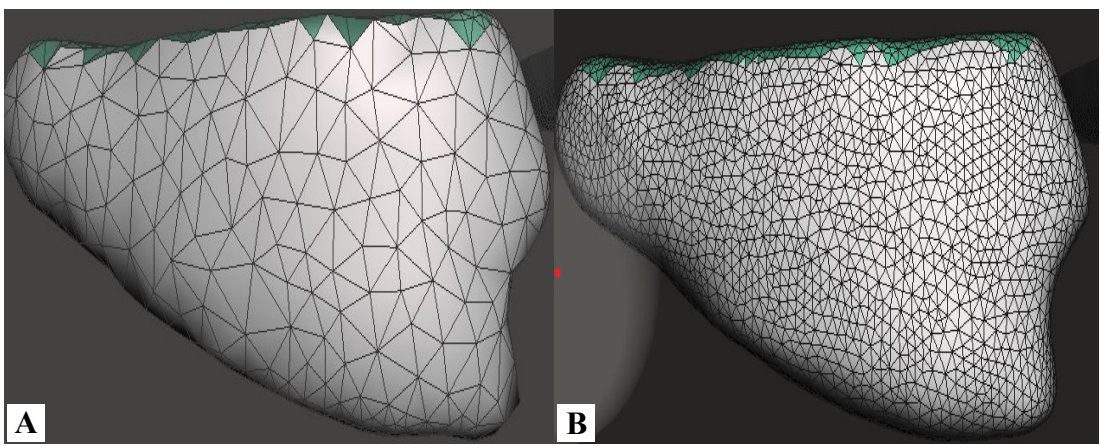

**Figure (5-15): Adjust the design (A) before edit surface (B) after edit surface**

After that, the jaw model adjusted at the beginning the is size reduced using the (Plan Cut) button from (Edit) menu. As presented in Figure (5-16). In Meshmixer, the holes in the model filled with (inspector) command from the (Analyze) menu as shown in Figure (5-17). Besides, sharp edges treated so that they are more suitable for transplantation.

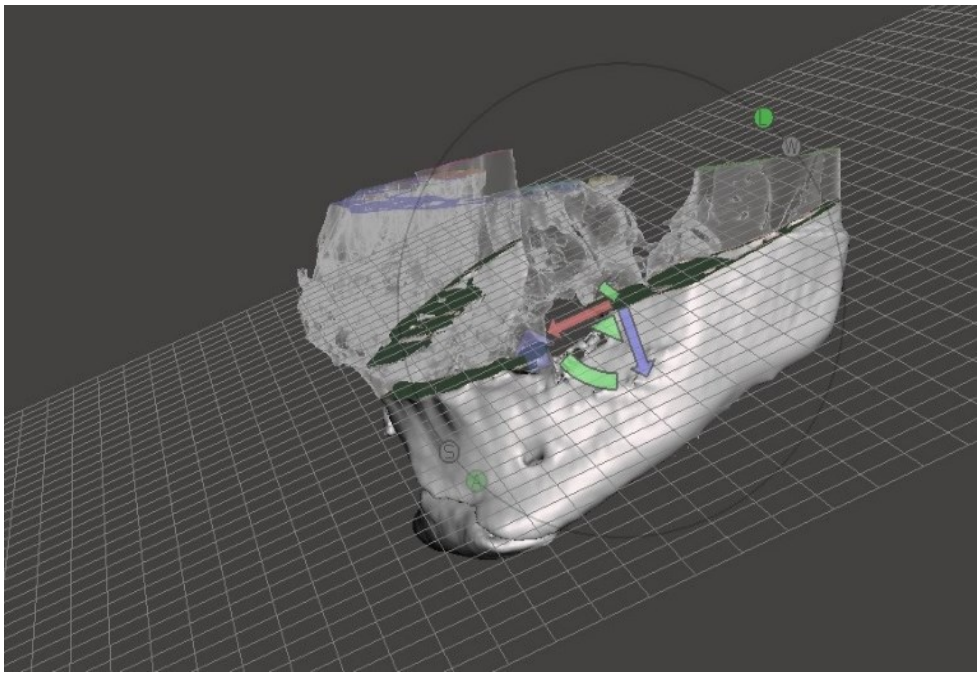

**Figure (5-16): Instructing (Plan Cut)**

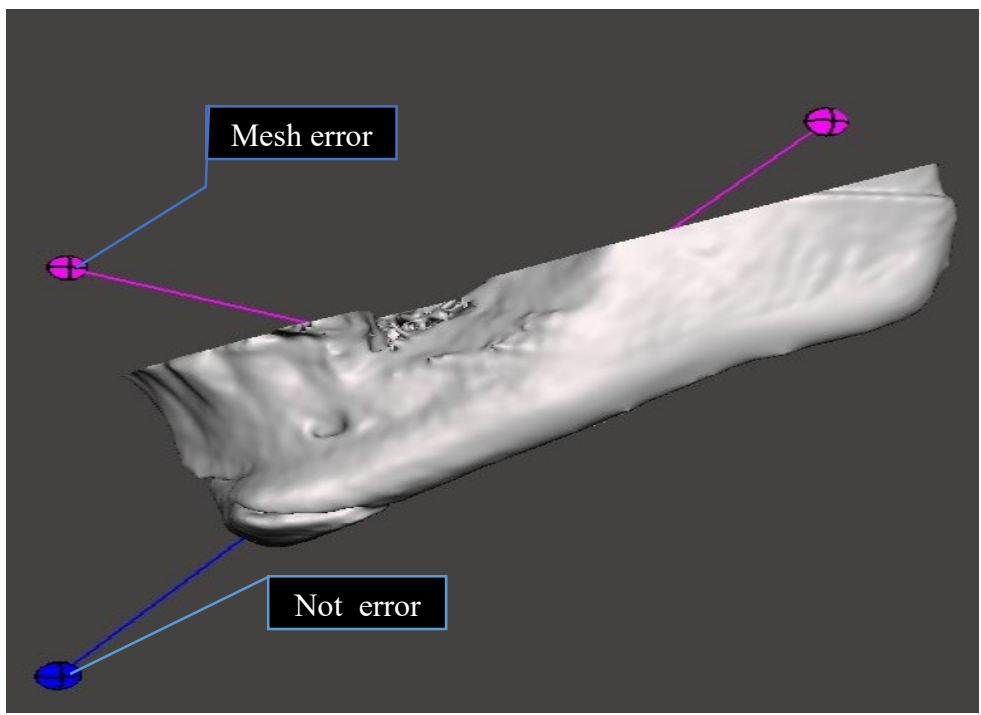

**Figure (5-17): Processing Of errors and holes**

Using the Transform technique, the broken part of the mandible repaired. Drives an angle of 11.45 degrees to the right to match the fraction with the mandibular, as shown in Figure (5-18). Figure (5-19) illustrates the transplantation using (Transform) technique. The transplant model moves from the platform area to the mandibular model. Uses drag and drop, the implant pulled into the mandibular cavity for conducting the matching process through the arrows. Also, dental openings are made.

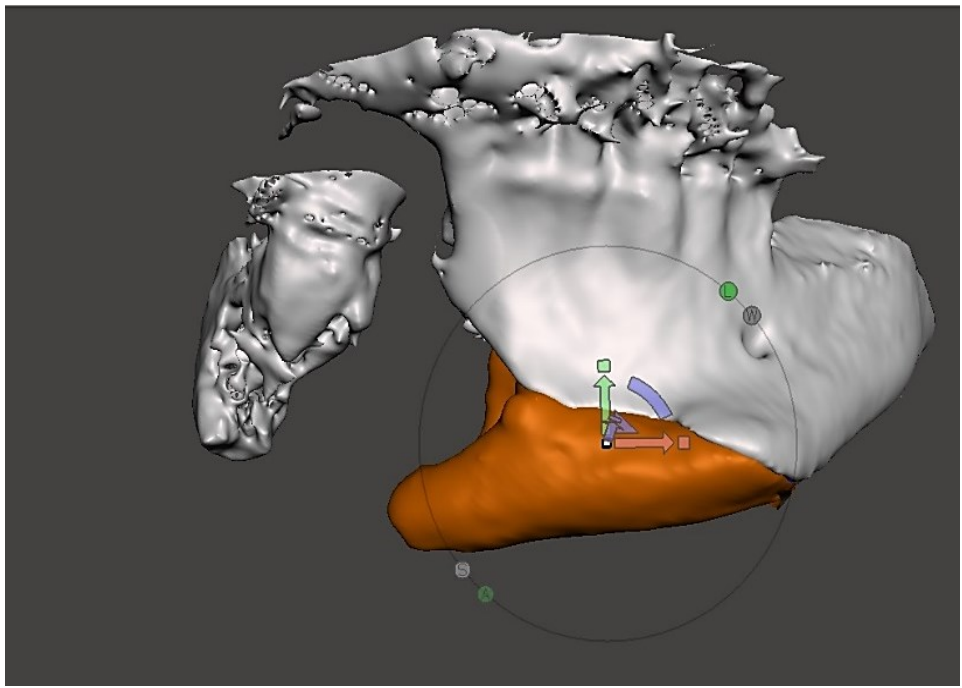

**Figure (5-18): Procedure Process Surgery of the mandibular fracture**

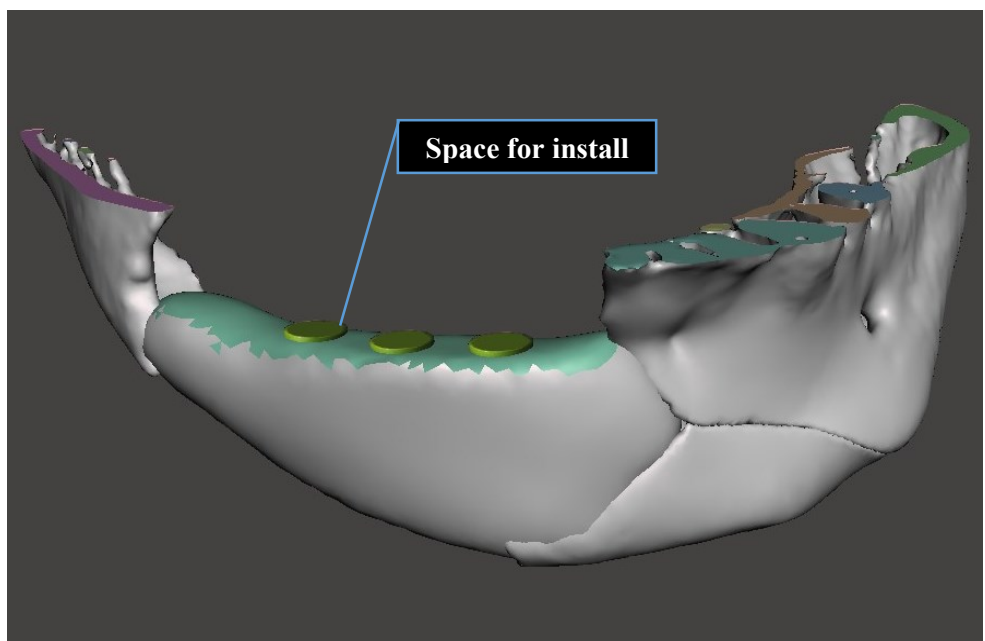

**Figure (5-19): The assembly process and the work of the tooth hole**

      Finally, the transplant is transformed into a reticle where some experienced doctors have been consulted about the transplant. Then select the "Make Pattern" See Figure (5-20). The command from the "Edit" menu is done then selecting the network type (Dual Edges) as well as specifying a value. The modification of the errors resulting from the process of switching to a network is done for improving the design as shown in Figure (5-21).

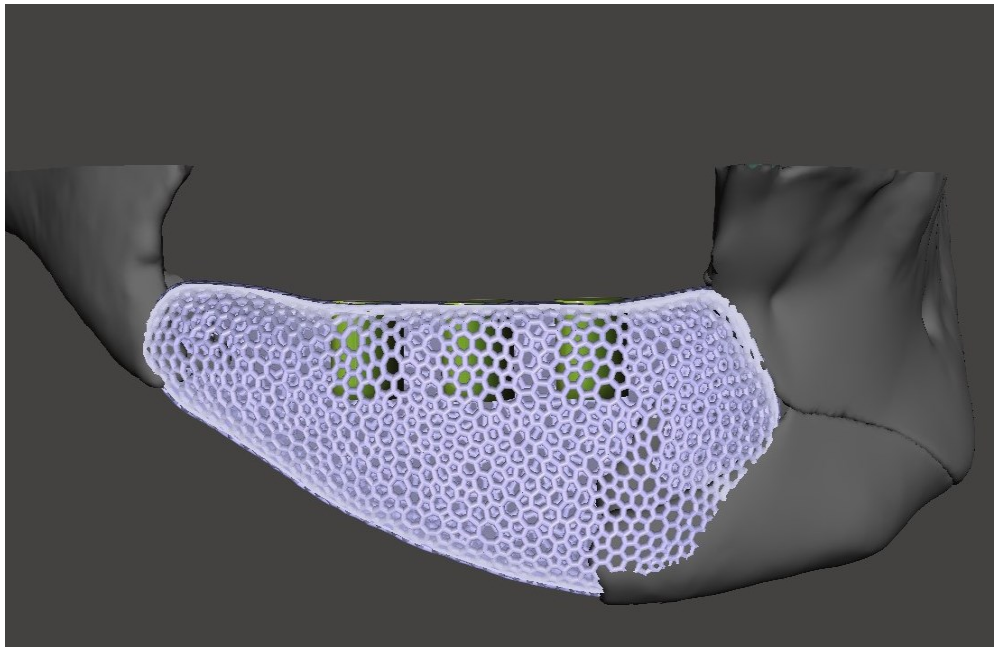

**Figure (5-20): Transform the implant into a mesh**

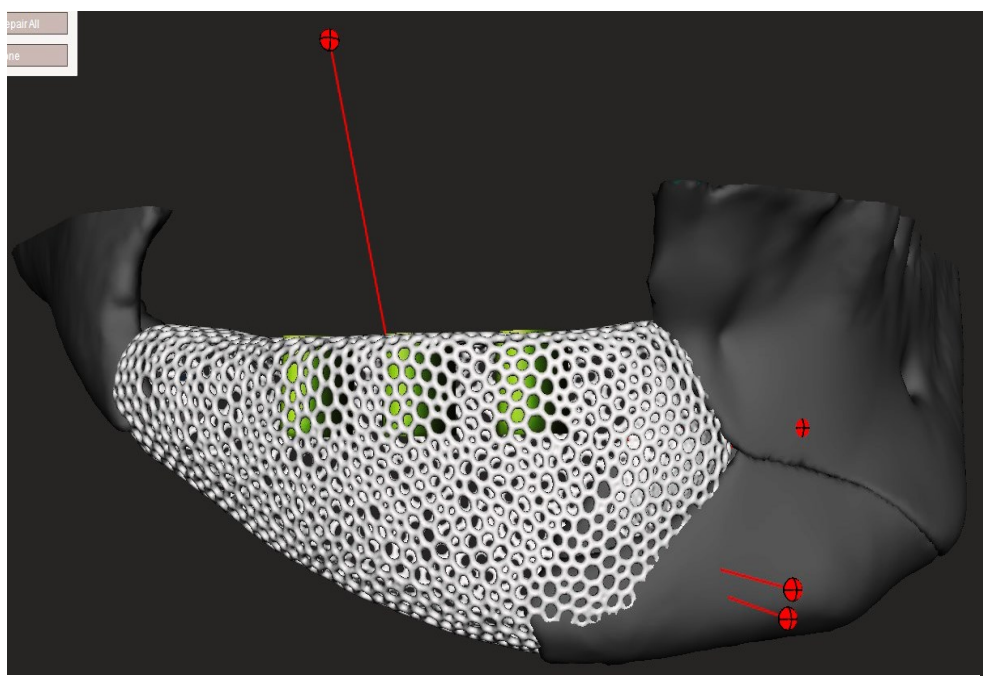

**Figure (5-21): Analysis and examination of holes**

Figure (5-22) illustrates the dimensions of the implant are length, width and height arranged respectively (41.7,16.6,19.6) mm. The Analysis menu is selected, select the Unit / Dimension command to determine the dimensions of the implant.

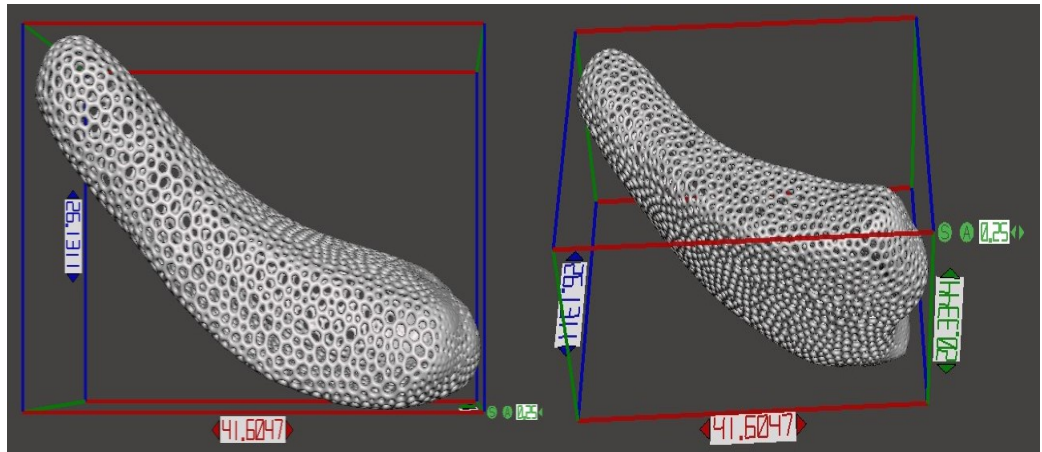

Figure (5-22): Dimensional Bounding Box

## 5.8 RP Medical Model Production

At this stage, the (SolidWork) will print a transplant form designed based on the patient's medical data, and modified by the (Meshmixer) program. The proposed method of fabrication in the methodology adopted is (SLS), but the lack its availability in Iraq prevented the achievement of this. The researcher has proposed an alternative fabrication method to represent the form and verify the model data (SLA).

### 5.8.1 Orientation

Once the part is considered a suitable building size, the part should be directed in an ideal position for construction. The shape of the part plays a key role in this case. The mandible and the dedicated implant should be properly oriented to avoid supporting material as much as possible as shown in Figure (5-23).

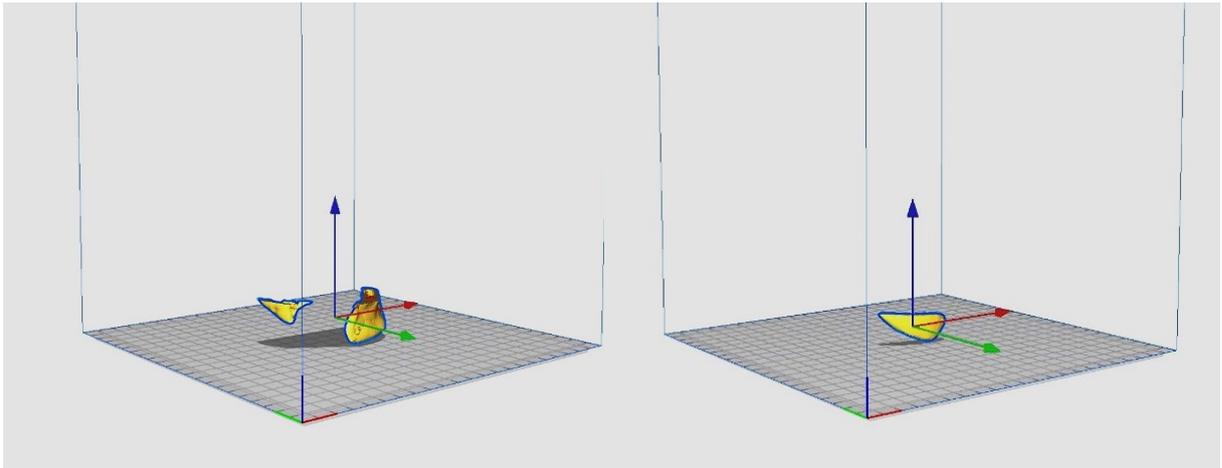

**Figure (5-23): STL mandible and implant imported into Cura software**

## 5.8.2 Slicing

Slicing is a software process that creates thin horizontal cross-sections of an STL file that will be used later to create the control code for the device as shown in Figure (5-24). In the Cura, the thickness of the slide can be changed before chipping, and typical slides range from 0.127mm to 3.81mm. Use thinner segments for high definition models.

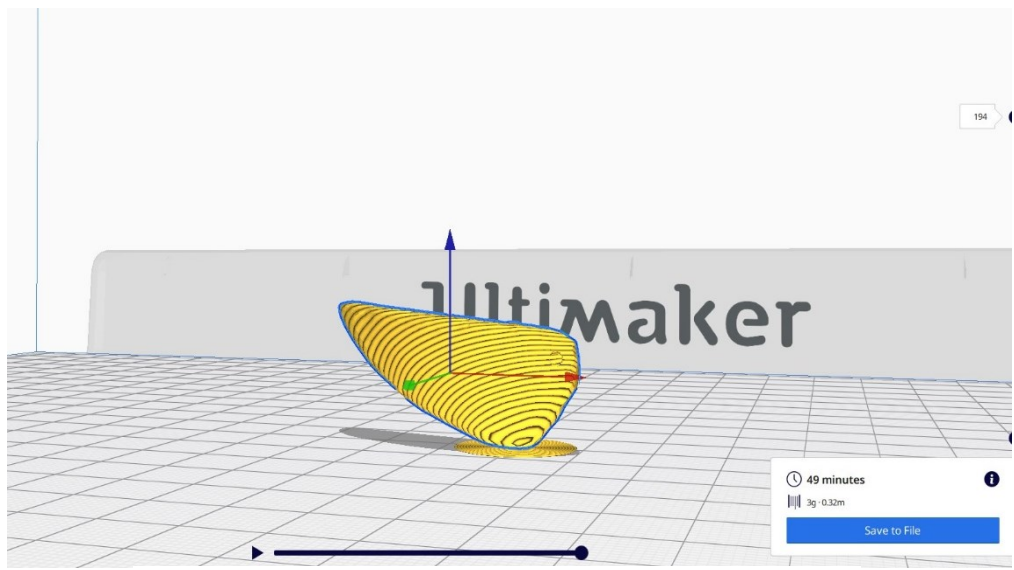

**Figure (5-24): Slicing the implant**

Built file saved on the flash memory card, which needs to be slot into Maker ware replicator.

## 5.8.3 Building a Part

SLA (ANYCUBIC PHOTON) technology works in a layered way as shown in Figure (5-25). When this process begins, the laser "pulls" the first layer of printing into a sensitive resin. Wherever the laser hits, the liquid hardens. The laser directed to the appropriate coordinates by a computer-controlled mirror. At this point, it should be not that most of the desktop SLA printers should work upside down. That is, the laser directed to the construction platform, which starts low and gradually raised. After the first layer, the platform aside according to the thickness of the layer (usually about 0.1 mm) and allows the additional resin to flow below the already printed part. The laser then hardens the next cross-section, and the process repeated until the entire part is completed. The laser untouched resin remains in the bowl and can be reused.

After polymerization of the material, the platform rises out of the tank and the excess resin discharged. At the end of the process, the model is removed from the platform; it washed from excess resin and then placed in a UV oven for final treatment. Generally, the post-print treatment enables objects to reach the highest possible strength and become more stable.

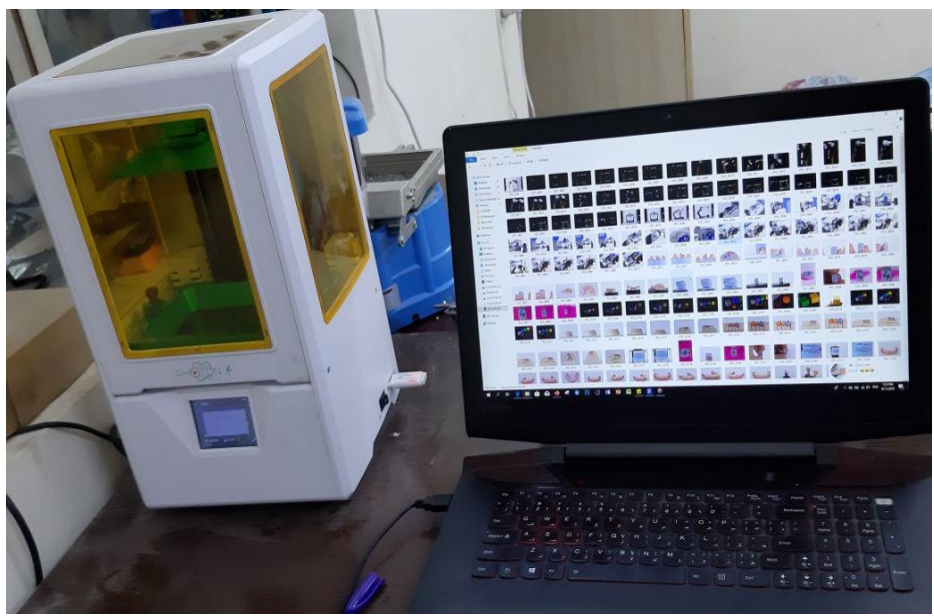

**Figure (5-25): SLA Printing**

## 5.9 Results and Discussion

The result shows a physical model of mandible produced during the current study has demonstrated that integration of RE/CAD/RP can be used successfully as shown in Figure (5-26).

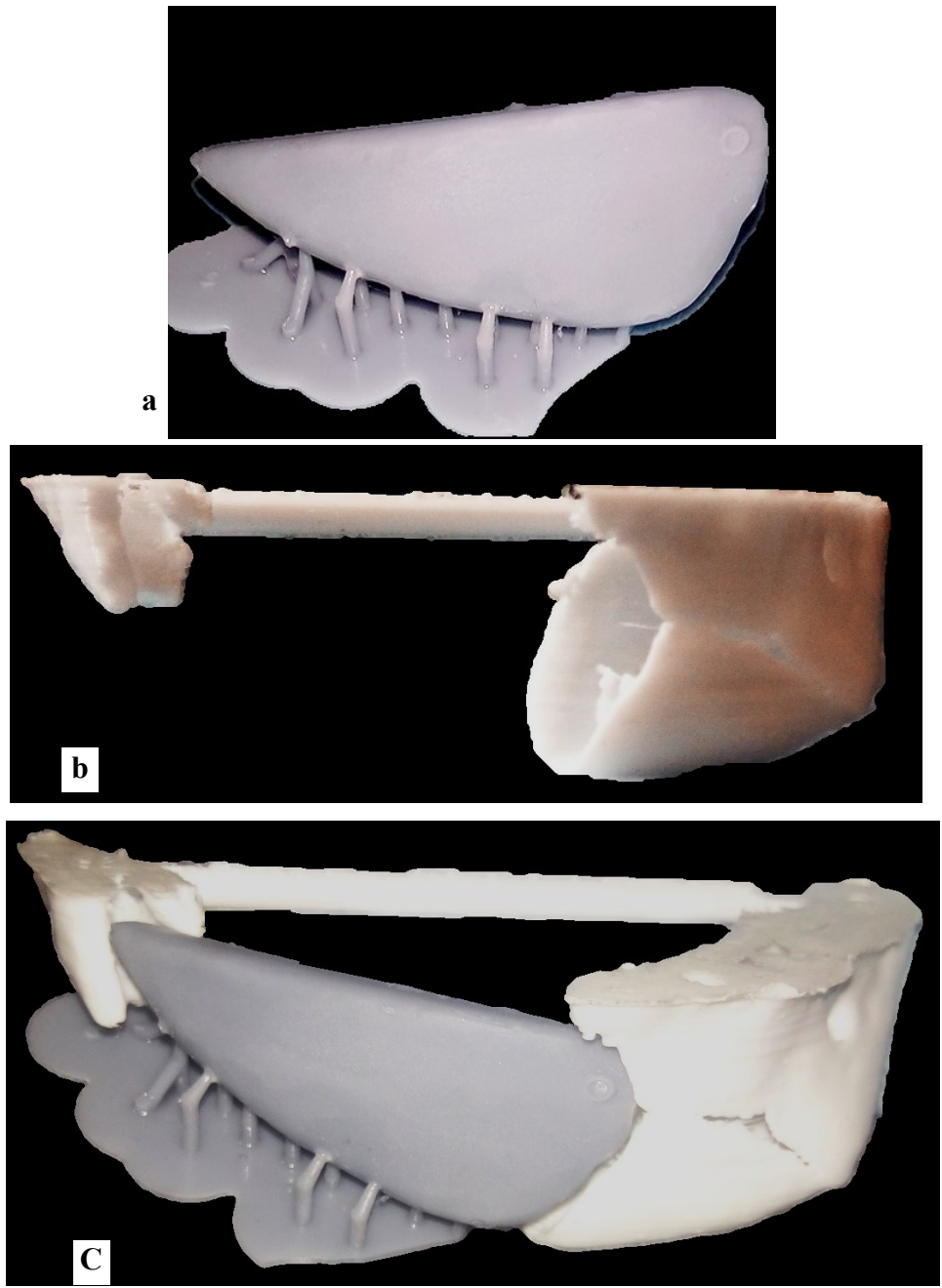

**Figure (5-26): (a): Implant model (missing part), (b): The mandible model, (C): Combining the two models (simulation)**

This work has been achieved through image data acquisition using a CBCT scan technic. The medical data was developed, refined and analyzed based on software (3DSlicer), thus obtain the prototype 3D model shown in Figure (b- 5-26). During the transformation of the model from the 2D to the 3D changes occur in the details of the model as a change in the thickness of the model and on this basis was used manual threshold technique. The medical model is designed within a SolidWorks environment. The curvature value depends on the skill and accuracy of the designer, based on data from the pathological condition.

The assembly of the models is carried out in an environment (Meshmixer) and the roughness of the extruded surfaces is treated due to the different noise resulting from the computerized tomography scan, and throughout the pre-treatment stage. There is a special tool, called "mesh smoothing" that can polish the residual peaks from noised roughness. Also, in this stage operation is performed surgical simulations and preoperative planning.

When creating layers for the printing process, the greater the number of layers, the greater the precision. While increasing the number of layers, it takes a long time to manufacture. The RP for this model took (49 min) and using (SLA), and the part was in the number of layers (194).

 In the final stage, it requires cleaning the part and removing the bolts from it. This process depends on the skill and accuracy of manual finishing operations that take some added time.

# CHAPTER SIX

# Conclusions and Recommendations

# CHAPTER SIX

## Conclusions and Recommendations

### 6.1 Conclusions

1. The methodology presented in this work can be applied in medical and several engineering applications.

2. Software (3DSlicer) can modeling, analyzing and modifying the represent the defected part.

3. The STL used for fabrication is segmented using surface rendering and then exported into the AM machine for the fabrication of customized implants.

4. A series of FEA procedures using Solidworks software is required for testing and validation of the mechanical properties of the finished implants.

5. The virtual computer environment that simulates clinical reality produces the successful anatomy of reconstructive surgery.

6. The quality of programs allows the surgeon to have a clear vision of the position in the surgical area and allowing to plan before surgery.

7. The integration method between the design and manufacturing environment provides a virtual computer environment that simulates the surgical clinical reality, thus improving surgical planning and reducing the time of the surgical operations.

### 6.2 Recommendations

1. It is suggested that the SLS technique using medical alloy (Ti6Al4V) should print the final model.

2. Dental implants can be added to the reconstructed mandible to improve the mandible functions, such as biting, chewing, and aesthetics aspects. That addition of dental implants improve the loading distribution on the mandible and reduce the risk of implant failure.

# Reference

[1] C. K. Chua, K. F. Leong, and J. An, "Introduction to rapid prototyping of biomaterials," *Rapid Prototyp. Biomater. Princ. Appl.*, pp. 1–15, 2014, Doi: 10.1533/9780857097217.1.

[2] A. Dehghanghadikolaei, N. Namdari, B. Mohammadian, and B. Fotovvati, "Additive Manufacturing Methods: A Brief Overview," *J. Sci. Eng. Res.*, vol. 5, no. 8, pp. 123–131, 2018.

[3] S. Nayar, S. Bhuminathan, and W. Bhat, "Rapid prototyping and stereolithography in dentistry," *J. Pharm. Bioallied Sci.*, vol. 7, no. 5, pp. 1–10, 2015, doi: 10.4103/0975-7406.155913.

[4] D. L. B. Joseph J. Beaman, Jr., Ming C. Leu & David W. Rosen, "A Brief History of Additive Manufacturing and the 2009 Roadmap for Additive Manufacturing: Looking Back and Looking Ahead," *RapidTech 2009 US-TURKEY Work. Rapid Technol.*, vol. 14, no. 1–10558, pp. 1–9, 2009.

[5] W. WANG, *REVERSE ENGINEERING TECHNOLOGY OF REINVENTION*. United States of America: CRC Press Taylor & Francis Group, 2011.

[6] S. Q. Bhumika Kapoor, Gaurav Singh, Rajendra Kumar Tewari, "Rapid prototyping: An innovative technique in dentistry Shakeba," *J. Oral Res. Rev.*, vol. 9, no. 2, pp. 96–102, 2017, doi: 10.4103/jorr.jorr.

[7] L. C. Hieu et al., "Medical rapid prototyping applications and methods," *Assem. Autom.*, vol. 25, no. 4, pp. 284–292, 2005, doi: 10.1108/01445150510626415.

[8] Francis O. Dompreh, "Application of Rapid Manufacturing Technologies to Integrated Product Development in Clinics and Medical Manufacturing Industries," *OhioLINK ETD Cent. Maag Libr. Circ. Desk*, p. 68, 2013.

[9] D. M. T. Dr. Ch. Vyshnavi, Dr. G. Kalpana & Dr. Aditya Sai Jagini, "CONCEPT OF RAPID PROTOTYPING AND ITS USES IN DENTISTRY,"

Int. J. Adv. Res., vol. 4, no. 9, pp. 953–957, 2016, doi: 10.21474/ijar01/1567.

[10] VENKAT KOLAR, "APPLICATION OF REVERSE ENGINEERING AND RAPID PROTOTYPING TO CASTING," OhioLINK ETD Cent. Maag Libr. Circ. Desk, pp. 1–68, 2004.

[11] W. P. Syam, M. A. Mannan, and A. M. Al-Ahmari, "Rapid prototyping and rapid manufacturing in medicine and dentistry," Virtual Phys. Prototyp., vol. 6, no. 2, pp. 79–109, 2011, doi: 10.1080/17452759.2011.590388.

[12] D. Hyndhavi and S. B. Murthy, "Rapid Prototyping Technology- Classification and Comparison," Int. Res. J. Eng. Technol., vol. 4, no. 6, pp. 3107–3111, 2017.

[13] V. Bagaria, D. Rasalkar, S. Jain, and J. Ilyas, "Medical Applications of Rapid Prototyping - A New Horizon," Adv. Appl. Rapid Prototyp. Technol. Mod. Eng., pp. 1–22, 2011, doi: 10.5772/20058.

[14] M. Stanek et al., "Rapid Prototyping Methods Comparison," Recent Res. Circuits Syst. Ind., pp. 269–272, 2012.

[15] M. Heller et al., "Applications of patient-specific 3D printing in medicine," Int. J. Comput. Dent., vol. 19, no. 4, pp. 323–339, 2016, doi: : https://www.researchgate.net/publication/316664541 Applications.

[16] M. P. Chae, W. M. Rozen, P. G. McMenamin, M. W. Findlay, R. T. Spychal, and D. J. Hunter-Smith, "Emerging Applications of Bedside 3D Printing in Plastic Surgery," Front. Surg., vol. 2, no. June, pp. 1–14, 2015, doi: 10.3389/fsurg.2015.00025.

[17] S. Negi, S. Dhiman, and R. K. Sharma, "Basics and applications of rapid prototyping medical models," Rapid Prototyp. J., vol. 20, no. 3, pp. 256–267, 2014, doi: 10.1108/RPJ-07-2012-0065.

[18] A. Manmadhachary, S. K. Malyala, and A. Alwala, "Medical applications of additive manufacturing," Lect. Notes Comput. Vis. Biomech., vol. 30, no.

*October, pp. 1643–1653, 2019, doi: 10.1007/978-3-030-00665-5_152.*

[19] *J. V. L. Silva, M. F. Gouvêia, A. Santa Barbara, E. Meurer, and C. A. C. Zavaglia, "Rapid Prototyping Applications in the Treatment of Craniomaxillofacial Deformities - Utilization of Bioceramics," Key Eng. Mater., vol. 254–256, pp. 687–690, 2004, doi: 10.4028/www.scientific.net/kem.254-256.687.*

[20] *J. Winder and R. Bibb, "Medical rapid prototyping technologies: State of the art and current limitations for application in oral and maxillofacial surgery," Am. Assoc. Oral Maxillofac. Surg., vol. 63, no. 7, pp. 1006–1015, 2005, doi: 10.1016/j.joms.2005.03.016.*

[21] *D. P. Sinn, J. E. Cillo, and B. A. Miles, "Stereolithography for craniofacial surgery," J. Craniofac. Surg., vol. 17, no. 5, pp. 869–875, 2006, doi: 10.1097/01.scs.0000230618.95012.1d.*

[22] *M. T. Jelena Milovanović, "Medical application of rapid prototyping," Med. Appl. Rapid Prototyp., vol. 5, no. 1, pp. 79–85, 2007, doi: 10.2493/jjspe.70.179.*

[23] *E. Kouhi, S. Masood, and Y. Morsi, "Design and fabrication of reconstructive mandibular models using fused deposition modeling," Assem. Autom., vol. 28, no. 3, pp. 246–254, 2008, doi: 10.1108/01445150810889501.*

[24] *S. Singare et al., "Rapid prototyping assisted surgery planning and custom implant design," Rapid Prototyp. J., vol. 15, no. 1355–2546, pp. 19–23, 2009, doi: 10.1108/13552540910925027.*

[25] *I. El-Katatny, S. H. Masood, and Y. S. Morsi, "Error analysis of FDM fabricated medical replicas," Rapid Prototyp. J., vol. 16, no. 1, pp. 36–43, 2009, doi: 10.1108/13552541011011695.*

[26] *L. Bin Zhou et al., "Accurate reconstruction of discontinuous mandible using a reverse engineering/computer-aided design/rapid prototyping technique: A*

*preliminary clinical study," J. Oral Maxillofac. Surg., vol. 68, no. 9, pp. 2115–2121, 2010, doi: 10.1016/j.joms.2009.09.033.*

[27] *E. Huotilainen et al., "Inaccuracies in additive manufactured medical skull models caused by the DICOM to STL conversion process," J. Cranio-Maxillofacial Surg., vol. 42, no. 5, pp. 1–7, 2013, doi: 10.1016/j.jcms.2013.10.001.*

[28] *E. A. Nasr, A. Al-Ahmari, A. Kamrani, and K. Moiduddin, "Digital design and fabrication of customized mandible implant," World Autom. Congr. Proc., pp. 1–6, 2014, doi: 10.1109/WAC.2014.6935880.*

[29] *M. A. Larosa, A. L. Jardini, C. A. D. C. Zavaglia, P. Kharmandayan, D. R. Calderoni, and R. Maciel Filho, "Microstructural and mechanical characterization of a custom-built implant manufactured in titanium alloy by direct metal laser sintering," Adv. Mech. Eng., vol. 2014, no. 945819, pp. 1–8, 2014, doi: 10.1155/2014/945819.*

[30] *A. L. Jardini et al., "Improvement in Cranioplasty: Advanced Prosthesis Biomanufacturing," Procedia CIRP, vol. 49, no. 2212–8271, pp. 203–208, 2016, doi: 10.1016/j.procir.2015.11.017.*

[31] *M. van Eijnatten, J. Koivisto, K. Karhu, T. Forouzanfar, and J. Wolff, "The impact of manual threshold selection in medical additive manufacturing," Int. J. Comput. Assist. Radiol. Surg., vol. 12, no. 1861–6410, pp. 607–615, 2016, doi: 10.1007/s11548-016-1490-4.*

[32] *A. M. Alwala1, L. R. C. and P. V. , Santosh Kumar Malyala2 *, A. Mohan Alwala, and S. Kumar Malyala, "Surgical Planning in Pan Facial Trauma Using Additive Manufacturing Medical Model-A Case Study," Jurnalul Chir., vol. 12, no. 3.8, pp. 125–128, 2016, doi: 10.7438/1584-9341-12-3-8.*

[33] *J. Egger et al., "Interactive reconstructions of cranial 3D implants under MeVisLab as an alternative to commercial planning software," PLoS One, vol.*

12, no. 3, pp. 1–20, 2017, doi: 10.1371/journal.pone.0172694.

[34] Z. Qin, Z. Zhang, X. Li, Y. Wang, P. Wang, and J. Li, "One-Stage treatment for maxillofacial asymmetry with orthognathic and contouring surgery using virtual surgical planning and 3D-printed surgical templates," J. Plast. Reconstr. Aesthetic Surg., vol. 72, no. 1, pp. 97–106, 2019, doi: 10.1016/j.bjps.2018.08.015.

[35] T. Ends, K. S. Daukanto, A. Maru, O. Raga, A. Paulauskas, and N. Tiso, "TRENDS IN PRODUCING PERSONALIZED BONE IMPLANTS USING ADDITIVE MANUFACTURING," in 13th INTERNATIONAL SCIENTIFIC CONFERENCE NOVI SAD, SERBIA, 2018, no. MMA 2018, pp. 371–374.

[36] T. F. Abbas, A. A. Khleif, and H. S. Ismael, "Design and Construction of a Device for Free Form Surfaces 3D Reconstruction Using Microsoft Kinect," Int. J. Sci. Res. Sci. Technol., vol. 6, no. 2, pp. 82–86, 2019, doi: 10.32628/ijsrst19629.

[37] M. Zhang, P. Rao, D. Xia, L. Sun, X. Cai, and J. Xiao, "Functional Reconstruction of Mandibular Segment Defects With Individual Preformed Reconstruction Plate and Computed Tomographic Angiography-Aided Iliac Crest Flap," J. Oral Maxillofac. Surg., vol. 77, no. 6, pp. 1293–1304, 2019, doi: 10.1016/j.joms.2019.01.017.

[38] I. Gibson et al., "The use of rapid prototyping to assist medical applications," Rapid Prototyp. J., vol. 12, no. 1, pp. 53–58, 2006, doi: 10.1108/13552540610637273.

[39] J. Han and Y. Jia, "CT image processing and medical rapid prototyping," in BioMedical Engineering and Informatics: New Development and the Future - Proceedings of the 1st International Conference on BioMedical Engineering and Informatics, 2008, vol. 2, pp. 67–71, doi: 10.1109/BMEI.2008.329.

[40] Z. L. Wei, Chao Zhang Mei, and P. S. Li, "Application of computer-aided three-

dimensional skull model with rapid prototyping technique in repair of zygomatico-orbito-maxillary complex fracture," Int. J. Med. Robot. Comput. Assist. Surg., no. 5, pp. 158–163, 2009, doi: 10.1002/rcs.

[41] J. M. Liliana Beldie, 1* Brian Walker, 1 Yongtao Lu, 2, 3 Stephen Richmond2, "Finite element modelling of maxillofacial surgery and facial expressions – a preliminary study," Int. J. Med. Robot. Comput. Assist. Surg., vol. 6, no. April, pp. 422–430, 2010, doi: 10.1002/rcs.

[42] P. U. Ilavarasi and M. Anburajan, "Design and finite element analysis of mandibular prosthesis," in 2011 3rd International Conference on Electronics Computer Technology (ICECT 2011), 2011, vol. 3, no. April, pp. 325–329, doi: 10.1109/ICECTECH.2011.5941765.

[43] K. V. Wong and A. Hernandez, "A Review of Additive Manufacturing," ISRN Mech. Eng., vol. 2012, pp. 1–10, 2012, doi: 10.5402/2012/208760.

[44] F. M. Othman, T. Abbas, and H. B. Ali, "Influence of Process Parameters on Mechanical Properties and Printing Time of FDM for PLA Printed Parts Using Design of Experiment.," Hind Basil Ali J. Eng. Res. Appl., vol. 8, no. 7, pp. 65–69, 2018, doi: 10.9790/9622-0807026569.

[45] M. Wang, X. Qu, M. Cao, D. Wang, and C. Zhang, "Biomechanical three-dimensional finite element analysis of prostheses retained with/without zygoma implants in maxillectomy patients," J. Biomech., vol. 46, no. 0021–9290, pp. 1155–1161, 2013, doi: 10.1016/j.jbiomech.2013.01.004.

[46] K. Salonitis, L. D'Alvise, B. Schoinochoritis, and D. Chantzis, "Additive manufacturing and post-processing simulation: laser cladding followed by high speed machining," Int. J. Adv. Manuf. Technol., vol. 85, no. 9–12, pp. 2401–2411, 2015, doi: 10.1007/s00170-015-7989-y.

[47] S. Singh, S. Ramakrishna, and R. Singh, "Material issues in additive manufacturing: A review," J. Manuf. Process., vol. 25, no. 1526–6125, pp.

185–200, 2017, doi: 10.1016/j.jmapro.2016.11.006.

[48] M. I. Mohammed, A. P. Fitzpatrick, and I. Gibson, "Customised design of a patient specific 3D printed whole mandible implant," KnE Eng., vol. 2, no. 2, p. 104, 2017, doi: 10.18502/keg.v2i2.602.

[49] S. Kumar Malyala, R. Y. Kumar, and A. M. Alwala, "A 3D-printed osseointegrated combined jaw and dental implant prosthesis – A case study," Rapid Prototyp. J., vol. 23, no. 6, pp. 1164–1169, 2017, doi: 10.1108/RPJ-10-2016-0166.

[50] J. C. Davies, H. H. L. Chan, Y. Jozaghi, D. P. Goldstein, and J. C. Irish, "Analysis of simulated mandibular reconstruction using a segmental mirroring technique," J. Cranio-Maxillofacial Surg., vol. 47, no. 3, pp. 468–472, 2018, doi: 10.1016/j.jcms.2018.12.016.

[51] A. Manmadhachary, S. K. Malyala, and A. Alwala, "Medical Applications of Additive Manufacturing," Springer Int. Publ., vol. 2018, no. January, pp. 1643–1653, 2018, doi: 10.1007/978-3-030-00665-5.

[52] I. Yavuz, M. F. Rizal, and B. Kiswanjaya, "The possible usability of three-dimensional cone beam computed dental tomography in dental research," J. Phys. Conf. Ser., vol. 884, no. 1, 2017, doi: 10.1088/1742-6596/884/1/012041.

## الخلاصة:

إعادة بناء الفك السفلي هي واحدة من التحديات الأكثر شيوعًا التي تواجه الجراحين. يلعب الفك السفلي دورًا رئيسيًا في دعم الأسنان داخل الفم.

يهدف هذا العمل إلى تطوير منهجية مقترحة لتحسين الجراحة الترميمية باستخدام عملية المحاكاة وبالاعتماد على تقنيات التصوير والتصميم والتصنيع للفك السفلي. يوفر الجمع بين هذه التقنيات طريقة قوية لتحسين وتنفيذ عملية الزرع من خلال تصميم وتصنيع النماذج الطبية.

يقدم هذا العمل منهجية تتمثل في إمكانات الهندسة العكسية (RE)، والتصميم بمساعدة الكمبيوتر (CAD)، وتقنية النماذج الأولية السريعة (RP) لإعادة إعمار الفك السفلي وتمثيل الجراحة. تم فحص المريض باستخدام تقنية عالية الدقة متمثلة في التصوير المقطعي المحوسب لمخروط الشعاع (CBCT)، يتم إنشاء نموذج رقمي تمثيلي للمريض ثلاثي الابعاد باستخدام برنامج (DSlicer3) للمساعدة في عملية التصميم. تم تصميم غرس المريض (جزء مفقود) وتحليله باستخدام برنامج Solidwork. يتم تجميع النماذج ثلاثية الأبعاد ويتم إجراء عمليات محاكاة الغرس بواسطة برنامج (Meshmixer).

تظهر نتائج هذا العمل أن إجراءات إعادة الإعمار الفك السفلي يمكن استخدامها بنجاح باستخدام تكامل تقنيات RE / CAD / RP. سيدعم هذا التكامل التماثل المقبول للوجه الذي سيتم استعادته من خلال مساعدة الجراحين في التخطيط لإعادة الإعمار. يوضح هذا العمل أهمية نظام CAD في حل متطلبات تلف المريض مباشرة من بيانات التصوير الطبي، وكذلك كفاءة تقنيات RP المستخدمة في تحويل تصاميم النماذج ثلاثية الأبعاد إلى نماذج زرع.

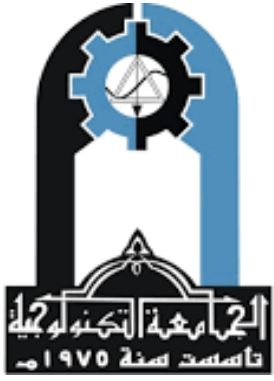

جمهورية العراق

وزارة التعليم العالي والبحث العلمي

الجامعة التكنولوجية

قسم هندسة الإنتاج والمعادن

# تطبيق الهندسة العكسية والنمذجة السريعة لإعادة بناء الفك السفلي البشري

رسالة مقدمة الى قسم هندسة الانتاج والمعادن /الجامعة التكنولوجية وهي جزء من متطلبات الدراسة نيل درجة الماجستير في علوم الهندسة الصناعية

اعداد

نبيل إسماعيل علاوي

بكالوريوس علوم في الهندسة الصناعية 2006

بأشراف

الدكتور أمجد بــــرزان عبــد الغــــفور

1441هـ

2020 م